\documentclass[structabstract]{aa}  
\usepackage{graphicx}
\usepackage{array}
\usepackage{txfonts}
\usepackage{natbib}
\bibpunct{(}{)}{;}{a}{}{,}
\usepackage{verbatim}
\usepackage{subfigure}
\usepackage{epsfig}

\newcommand{\vy}[2]{#1_{\scriptscriptstyle #2}}

\def\gtorder{\mathrel{\raise.3ex\hbox{$>$}\mkern-14mu
             \lower0.6ex\hbox{$\sim$}}}
\def\ltorder{\mathrel{\raise.3ex\hbox{$<$}\mkern-14mu
             \lower0.6ex\hbox{$\sim$}}}
\def\proptwid{\mathrel{\raise.3ex\hbox{$\propto$}\mkern-14mu
             \lower0.6ex\hbox{$\sim$}}}
\def\0946{PG~0946+301}

\def\arcsec{\ifmmode '' \else $''$\fi}

\def\arcsecpoint{\ifmmode ''\!. \else $''\!.$\fi}

\def\kms{\ifmmode {\rm km\ s}^{-1} \else km s$^{-1}$\fi}
\def\Msun{\ifmmode {\rm M}_{\odot} \else M$_{\odot}$\fi}
\def\Lsun{\ifmmode {\rm L}_{\odot} \else L$_{\odot}$\fi}
\def\Zsun{\ifmmode {\rm Z}_{\odot} \else Z$_{\odot}$\fi}

\def\ergscm2{ergs\,s$^{-1}$\,cm$^{-2}$}
\def\icm3{{\rm cm}^{-3}}
\def\icm2{{\rm cm}^{-2}}
\def\qo{\ifmmode q_{\rm o} \else $q_{\rm o}$\fi}
\def\Ho{\ifmmode H_{\rm o} \else $H_{\rm o}$\fi}
\def\ho{\ifmmode h_{\rm o} \else $h_{\rm o}$\fi}
\def\ltsim{\raisebox{-.5ex}{$\;\stackrel{<}{\sim}\;$}}
\def\gtsim{\raisebox{-.5ex}{$\;\stackrel{>}{\sim}\;$}}
\def\vFWHM{\ifmmode v_{\mbox{\tiny FWHM}} \else
            $v_{\mbox{\tiny FWHM}}$\fi}
\def\CCF{\ifmmode F_{\it CCF} \else $F_{\it CCF}$\fi}
\def\ACF{\ifmmode F_{\it ACF} \else $F_{\it ACF}$\fi}
\def\Halpha{\ifmmode {\rm H}\alpha \else H$\alpha$\fi}
\def\Hbeta{\ifmmode {\rm H}\beta \else H$\beta$\fi}
\def\Hgamma{\ifmmode {\rm H}\gamma \else H$\gamma$\fi}
\def\Hdelta{\ifmmode {\rm H}\delta \else H$\delta$\fi}
\def\Lya{\ifmmode {\rm Ly}\alpha \else Ly$\alpha$\fi}
\def\Lyb{\ifmmode {\rm Ly}\beta \else Ly$\beta$\fi}
\def\Lyg{\ifmmode {\rm Ly}\beta \else Ly$\gamma$\fi}

\def\hii{H\,{\sc ii}}

\def\ciii{\ifmmode {\rm C}\,{\sc iii} \else C\,{\sc iii}\fi}
\def\civ{\ifmmode {\rm C}\,{\sc iv} \else C\,{\sc iv}\fi}

\def\nv{N\,{\sc v}}

\def\o5007{[O\,{\sc iii}]\,$\lambda5007$}

\def\siiv{Si\,{\sc iv}}
\def\siIV{Si\,{\sc iv}}

\def\o{\o}
\begin{document}

   \title{Multiwavelength Campaign on Mrk 509}
   \subtitle{X. Lower limit on the distance of the absorber from HST COS and
STIS spectroscopy}

\author{N. Arav\inst{1}
  \and D. Edmonds\inst{1}
  \and B. Borguet\inst{1}
  \and G.A. Kriss\inst{2,3}
  \and J.S. Kaastra\inst{4,5}
  \and E. Behar\inst{6}
  \and S. Bianchi\inst{7}
  \and M. Cappi\inst{8}
  \and E. Costantini\inst{4}
  \and R.G. Detmers\inst{4,5}
  \and J. Ebrero\inst{4}
  \and M. Mehdipour\inst{9}
  \and S. Paltani\inst{10}
  \and P.O. Petrucci\inst{11}
  \and C. Pinto\inst{4}
  \and G. Ponti\inst{12}
  \and K.C. Steenbrugge\inst{13,14}
  \and C.P. de Vries\inst{4}
  }
  
\institute{Department of Physics, Virginia Tech, Blacksburg, Va 24061; email:
arav@vt.edu 
	   \and Space Telescope Science Institue, 3700 San Martin Drive,
Baltimore, MD 21218
	   \and Department of Physics \& Astronomy, The Johns Hopkins
University, Baltimore, MD 21218
	   \and SRON Netherlands Institute for Space Research, Sorbonnelaan 2,
3584 CA Utrecht, the Netherlands
	   \and Sterrenkundig Instituut, Universiteit Utrecht, P.O. Box 80000,
3508 TA Utrecht, the Netherlands
Holmsbury St. Mary, Dorking, Surrey, RH5 6NT, UK
	   \and Department of Physics, Technion-Israel Institute of Technology,
Haifa 32000, Israel
           \and Dipartimento di Fisica, Universita degli Studi Roma Tre, via
della Vasca Navale 84, 00146 Roma, Italy
	   \and INAF-IASF Bologna, Via Gobetti 101, 40129 Bologna, Italy
           \and Mullard Space Science Laboratory, University College London,
Holmsbury St. Mary, Dorking, Surrey, RH5 6NT, UK
	   \and ISDC Data Centre for Astrophysics, Astronomical Observatory of
the University of Geneva, 16 ch. d'Ecogia, 1290 Versoix, Switzerland
	   \and UJF-Grenoble 1 / CNRS-INSU, Institut de Planetologie et
d'Astrophysique de Grenoble (IPAG) UMR 5274, Grenoble, F-38041, France
	   \and School of Physics and Astronomy, University of Southampton,
Highfield, Southampton SO17 1BJ
	   \and Instituto de Astronomia, Universidad Catolica del Norte, Avenida
Angamos 0610, Casilla 1280, Antofagasta, Chile
	   \and Department of Physics, University of Oxford, Keble Road, Oxford
OX1 3RH, UK
} 
\date{\today}

\abstract
{}
{Active Galactic Nuclei often show evidence of photoionized outflows. A major
uncertainty in models for these outflows is the distance ($R$) to the gas from
the central black hole. In this paper we use the HST/COS data from a massive
multi-wavelength monitoring campaign on the bright Seyfert I galaxy Mrk 509, in
combination with archival HST/STIS data, to constrain the location of the
various kinematic components of the outflow.}
{We compare the expected response of the photoionized gas to
changes in ionizing flux with the changes measured in the data using the
following steps:
1) We compare the column densities of each kinematic component measured in the
2001 STIS data with those measured in the 2009 COS data; 2) We use
time-dependent photionization calculations with a set of simulated
lightcurves to put statistical upper limits on the hydrogen number density
($\vy{n}{H}$) that are consistent with the observed small changes in the ionic
column densities; 3) From the upper limit on $\vy{n}{H}$, we calculate a lower
limit on the distance to the absorber from the central source via the prior
determination of the ionization parameter. Our method offers two
improvements on traditional timescale analysis. First, we account for the
physical behavior of AGN lightcurves. Second, our analysis accounts for the quality
of measurement in cases where no changes are observed in the absorption
troughs.}
{The very small variations in trough ionic column densities (mostly consistent
with no change) between the 2001 and 2009 epochs allow us to put
statistical lower limits on $R$ between 100--200 pc for all the major UV
absorption components at a confidence level of 99\%. These results are mainly
consistent with the independent distance estimates derived for the warm
absorbers from the simultaneous X-ray spectra. Based on the 100--200 pc
lower limit for all the UV components, this
absorber cannot be connected with an accretion disc wind. The outflow might have
originated from the disc, but based on simple ballistic  kinematics, such an
event had to occur at least 300,000 years ago in the rest frame of the source.}
{}

\maketitle

\section{Introduction}

Outflows from active galactic nuclei (AGN) are detected as blue-shifted spectral
absorption features with respect to the rest frame of the AGN (e.g.,
\citealt{Crenshaw00,Crenshaw03,Kriss00,Arav02}).  Measurements of the absorption
troughs, combined with photoionization modeling,  yield the ionization parameter
($U_H$) and total column density of the gas  ($N_H$). However spectral data do
not provide a direct measurement for the distance ($R$) to the outflow from the
central source, and most outflows are unresolved point sources on images.
Therefore, we use indirect methods to obtain $R$, where the most common ones use
the relationship $U_H\propto(\vy{n}{H}R^2)^{-1}$. Since $U_H$ can be determined
from photoionization modeling, knowledge of the hydrogen number density
($\vy{n}{H}$) yields $R$.

In spectra where absorption features due to excited states of a given ion are
detected, the ratio of column densities from excited and ground levels can yield
$\vy{n}{H}$ (e.g.,
\citealt{deKool01,Hamann01,Korista08,Moe09,Dunn10a,Edmonds11}). Alternatively,
determining how the absorber responds to changes in the ionizing flux can
provide reliable estimates of $\vy{n}{H}$. Time-variability of the continuum is
a known feature of AGN (e.g., \citealt{Uttley03,McHardy06,Ishibashi09}). How the
fractional population of each ion in the outflowing gas changes in response to
variation in the ionizing continuum depends on the electron number density
$n_e$, which is $\simeq 1.2\vy{n}{H}$ in highly ionized gas. The time in which
the absorber adjusts to the new flux level is inversely proportional to $n_e$.
Therefore, by tracking changes in ionic column densities and ionizing flux over
time we can estimate $\vy{n}{H}$ and thus, the distance (e.g.
\citealt{Nicastro99,Gabel05b}).

As part of a large multiwavelength campaign \citep[hereafter Paper
I]{Kaastra11a}, the bright Seyfert I galaxy Mrk 509 was observed with the Cosmic
Origins Spectrograph (COS) onboard the \textit{Hubble} Space Telescope (for details, see
\citealt[hereafter Paper VI]{Kriss11}). Since the Mrk 509 UV spectra do not
contain troughs from excited states, we use time-variability to constrain the
distance to the absorber. This is done by: 1) comparing the column densities of
each kinematic component measured in the 2001 STIS data \citep{Kraemer03} to
those measured in the 2009 COS data (Paper VI); 2) determining the upper limit
on $\vy{n}{H}$ that is consistent with the small observed changes in the ionic
column densities. From the upper limit on  $\vy{n}{H}$, we then calculate the
lower limit on $R$ via the determined $U_H$. This method has been applied to the
Mrk 509 X-ray data by \citet[hereafter Paper VIII]{Kaastra12} using the
observed lightcurve over a 100 day monitoring campaign. Since the lightcurve
for Mrk 509 was was not monitored between the the 2001 STIS epoch and the start
of our campaign, we use Monte Carlo simulations to develop a sample of
lightcurves to put statistical limits on $\vy{n}{H}$. In section 6, we
will show the limits derived from timescale arguments are similar to and bracket
our statistical limits.

The paper is structured as follows. In section 2, we discuss the measurements of
and changes in absorption troughs between the 2001 and 2009 epochs.
Photoionization solutions are given in section 3, and time-dependent ionization
is discussed in section 4. The simulations used to determine $\vy{n}{H}$ are
discussed in section 5 along with distance determinations. We discuss our
results in section 6. In the appendix, we provide an illustrative example of the
time-dependent photoionization equations for the case of hydrogen, which can be
solved analytically.

\section{Comparison of UV Spectra from the 2001 and 2009 Epochs}

To quantitatively establish differences in UV absorption between the epochs of
the STIS observation (2001 April 13) and
the COS observations (2009 December 10 and 11),
we use the calibrated COS and STIS spectra presented in Paper VI.
For STIS, the spectrum is the one-dimensional archival echelle spectrum
originally obtained by Kraemer et al. (2003), re-reduced with up-to-date
pipeline
processing that includes corrections for scattered light and echelle blaze
evolution.
In Paper VI, we discuss custom calibrations for the COS spectrum that include
improved
wavelength calibrations (accurate to 5 $\rm km~s^{-1}$), flat field, and
flux calibrations. In addition, the COS spectrum was deconvolved (see Paper VI)
to correct for the broad wings of the line-spread
function in COS (Ghavamian et al. 2009; Kriss 2011).
As we showed in Paper VI, this deconvolution is important for an
accurate comparison between the COS and STIS spectra.
Comparison of the depths and widths of interstellar lines that are common to
the COS and STIS spectra (e.g., Figure 3 in Paper VI)
give us confidence that the deconvolved COS spectrum
is an accurate measure of the spectral properties of Mrk 509---saturated ISM
lines in the COS spectrum are black and have the same width as in the STIS
spectrum, and unsaturated lines have the same depth and width.

\begin{figure}
   \includegraphics[width=8.5cm, trim=72 72 72 72]{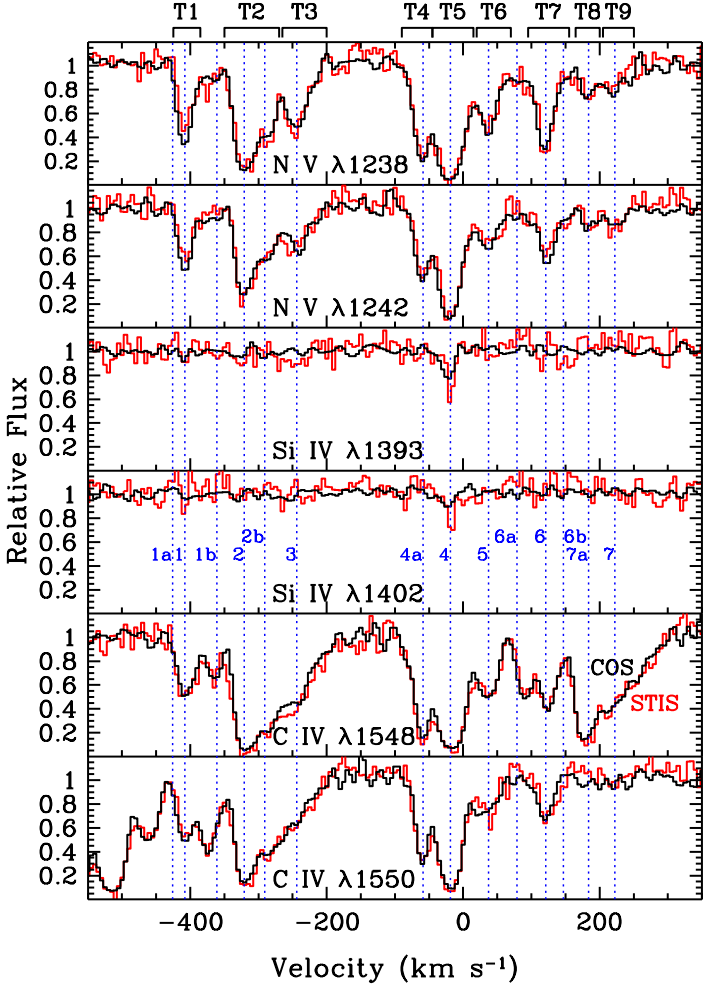}
  \caption{Comparison of spectral features in the COS (black) and
STIS (red) spectra of Mrk 509.
Normalized relative fluxes are plotted as a function of velocity relative to
the systemic redshift of $z=0.034397$.
The boundaries of the 9 absorption troughs used in our analysis are shown
along the top (see Table~\ref{coldensi}). The centroids of the individual
Gaussian components identified in Paper VI are shown by the vertical dotted blue
lines.
}
  \label{fig_cosstis}
\end{figure}

The calibrated spectra were divided by the best fitting emission model, which
includes continuum and emission lines (presented in Paper VI).
The same emission model was used for both STIS and COS spectra, but with
appropriately
fitted adjustments to the intensities of the emission lines and
the continuum.
We then rebinned these spectra onto the identical
velocity scale using velocity bins of 5 $\rm km~s^{-1}$.
This scale combines multiple pixels in each bin for both STIS and COS spectra,
reducing the
correlated errors introduced when dividing original pixels between adjacent
velocity bins.
This scale also gives about 3 bins per COS resolution element, so we preserve
the full resolution of the COS spectrum.
The Ly$\alpha$ absorption trough in Mrk 509 is heavily saturated, with
differences in absorption dominated by differences in covering fraction
rather than optical depth. We therefore confine our analysis to the unsaturated
\nv, \siiv, and \civ\ absorption lines.
Fig. \ref{fig_cosstis} compares the STIS and COS absorption troughs
for each transition of \nv, \siiv, and \civ.
We note that due to the close spacing of the  \civ\
$\lambda\lambda1548,1550$ doublet, the troughs of these transitions overlap
in velocity in the ranges $-498$ to $-375~\rm km~s^{-1}$ and
$+75$ to $+260~\rm km~s^{-1}$.
In Paper VI, the absorption in the COS spectrum was fit using 14 Gaussian
components. For comparing column densities, we use the \nv\ profiles to define 9
independent
troughs with velocity boundaries as given in Table~\ref{coldensi}.

There was an overall increase in flux for the COS spectrum compared to the
STIS spectrum, with an average increase of 72\% (see Paper VI for details). For
comparison to the historical lightcurve for Mrk 509, at rest wavelength
1354~$\AA$, $F_\lambda $(STIS)$ = 0.849 \times
10^{-13}$~erg~cm$^{-2}$~s$^{-1}$~$\AA^{-1}$, $F_\lambda $(COS)$ = 1.46 \times
10^{-13}$~erg~cm$^{-2}$~s$^{-1}$~$\AA^{-1}$, and the mean flux of the lightcurve
given in Section 5, Figure 3 is $F_\lambda $(mean)$ = 0.704 \times
10^{-13}$~erg~cm$^{-2}$~s$^{-1}$~$\AA^{-1}$.

\subsection{absorption trough variability}

To make a quantitative comparison between the COS and STIS spectra, for each
trough in each spectrum we calculate the mean transmission
\begin{equation}
\left< T \right> = {{\sum T_i} \over {N}},
\end{equation}
where $ T_i$ is the transmission in velocity bin $i$, and $N$ is the total
number of bins in a trough.
We also calculate the mean observed difference in transmission between the
COS and STIS spectra, $\left< T_C - T_S \right>$, and the mean error in this
difference,
\begin{equation}
\left< \sigma \right> = {{\sqrt {\sum (\sigma_{i,COS}^2 + \sigma_{i,STIS}^2)}}
\over {N}}.
\end{equation}
Due to the high S/N of the COS data, the error in the difference is dominated by
the
statistical errors in the STIS spectrum.
Table~\ref{coldensi} gives the mean transmissions of each trough as
observed in the COS spectrum, the mean fractional difference between the
COS and STIS troughs normalized by the mean COS transmission, and the mean
fractional error in this difference, again normalized by the
mean COS transmission.

As one can see in Figure~\ref{fig_cosstis} and Table~\ref{coldensi}, the
absorption in Mrk 509 showed little variation between the 2001 STIS spectrum
and the 2009 COS spectrum.
Our criterion for a significant variation requires that
both the red and the blue components of a trough show a difference of
$> 2\sigma$. In Table~\ref{coldensi}, this means that the absolute value in the
last column is greater than 2.
In Paper VI, we noted a significant difference in the \nv\ absorption
in trough T1, and that is apparent in the comparison shown in
Table~\ref{coldensi}.
Both the red and blue doublets of \nv\ show more than a $2 \sigma$
difference in transmission between the COS and STIS spectra.
However, no other trough in \nv\ or \siiv\ meets this criterion,
and only trough T2 in \civ\ shows such a significant difference.

\begin{table*}[!tbp]
\caption{variability in mrk 509 absorption troughs} 
\centering                
\begin{tabular}{cccccccc} 
\hline\hline
Feature & Trough & $v_1$ & $v_2$ & $ \left< T_C \right> $\tablefootmark{a} &
$\left< T_C-T_S \right> / \left< T_C \right>$\tablefootmark{b} & $\left<
\sigma\right> / \left< T_C \right>$\tablefootmark{c} &  $\left< T_C-T_S \right>
/ \left< \sigma \right> $\tablefootmark{d} \\
\hline
\nv\ $\lambda$ 1238 &  T1 & -425   &  -385 &      0.605 &     -0.182 &     0.039
&     -4.7 \\
\nv\ $\lambda$ 1238 &  T2 & -350   &  -270 &      0.425 &      0.004 &     0.035
&      0.1 \\
\nv\ $\lambda$ 1238 &  T3 & -265   &  -210 &      0.669 &      0.030 &     0.031
&      1.0 \\
\nv\ $\lambda$ 1238 &  T4 &  -90.0 & -45.0 &      0.530 &     -0.044 &     0.042
&     -1.1 \\
\nv\ $\lambda$ 1238 &  T5 &  -40.0 &  15.0 &      0.249 &     -0.092 &     0.065
&     -1.4 \\
\nv\ $\lambda$ 1238 &  T6 &   20.0 &  70.0 &      0.664 &     -0.023 &     0.030
&     -0.8 \\
\nv\ $\lambda$ 1238 &  T7 &   95.0 &   155 &      0.647 &     -0.005 &     0.028
&     -0.2 \\
\nv\ $\lambda$ 1238 &  T8 &    165 &   200 &      0.825 &      0.020 &     0.030
&      0.6 \\
\nv\ $\lambda$ 1238 &  T9 &    205 &   250 &      0.831 &     -0.079 &     0.027
&     -3.0 \\
\hline
\nv\ $\lambda$ 1242 &  T1 & -425   &  -385 &      0.715 &     -0.092 &     0.030
&     -3.0 \\
\nv\ $\lambda$ 1242 &  T2 & -350   &  -270 &      0.590 &      0.013 &     0.023
&      0.5 \\
\nv\ $\lambda$ 1242 &  T3 & -265   &  -210 &      0.774 &      0.008 &     0.024
&      0.3 \\
\nv\ $\lambda$ 1242 &  T4 &  -90.0 & -45.0 &      0.688 &     -0.052 &     0.027
&     -1.9 \\
\nv\ $\lambda$ 1242 &  T5 &  -40.0 &  15.0 &      0.376 &     -0.096 &     0.039
&     -2.4 \\
\nv\ $\lambda$ 1242 &  T6 &   20.0 &  70.0 &      0.798 &     -0.034 &     0.024
&     -1.4 \\
\nv\ $\lambda$ 1242 &  T7 &   95.0 &   155 &      0.774 &     -0.049 &     0.023
&     -2.2 \\
\nv\ $\lambda$ 1242 &  T8 &    165 &   200 &      0.901 &     -0.027 &     0.026
&     -1.0 \\
\nv\ $\lambda$ 1242 &  T9 &    205 &   250 &      0.916 &      0.003 &     0.022
&      0.1 \\
\hline
\siIV\ $\lambda$ 1393 & T5 & -45.0 &  15.0 &     0.937  &      0.006 &    0.023 
&     0.3 \\
\siIV\ $\lambda$ 1402 & T5 & -45.0 &  15.0 &     0.987  &      0.060 &    0.021 
&     2.9 \\
\hline
\civ\ $\lambda$ 1548 &  T1 & -425  &  -395 &      0.643 &      0.060 &     0.034
&       1.8 \\
\civ\ $\lambda$ 1548 &  T2 & -350  &  -270 &      0.294 &      0.190 &     0.035
&       5.5 \\
\civ\ $\lambda$ 1548 &  T3 & -265  &  -200 &      0.613 &      0.010 &     0.023
&       4.3 \\
\civ\ $\lambda$ 1548 &  T4 & -90.0 & -45.0 &      0.475 &      0.101 &     0.030
&       3.4 \\
\civ\ $\lambda$ 1548 &  T5 & -40.0 &  15.0 &      0.261 &      0.120 &     0.041
&       2.9 \\
\civ\ $\lambda$ 1548 &  T6 &  20.0 &  70.0 &      0.694 &     -0.007 &     0.024
&     -0.3 \\
\civ\ $\lambda$ 1548 &  T7 &  95.0 &   155 &      0.601 &      0.017 &     0.023
&      0.8 \\
\civ\ $\lambda$ 1548 &  T8 &   165 &   200 &      0.246 &      0.118 &     0.047
&       2.5 \\
\civ\ $\lambda$ 1548 &  T9 &   205 &   250 &      0.486 &      0.009 &     0.029
&      0.3 \\
\hline
\civ\ $\lambda$ 1550 &  T1 & -425  &  -395 &      0.562 &     -0.010 &     0.033
&     -0.3 \\
\civ\ $\lambda$ 1550 &  T2 & -350  &  -270 &      0.404 &      0.070 &     0.024
&       2.9 \\
\civ\ $\lambda$ 1550 &  T3 & -265  &  -200 &      0.749 &      0.026 &     0.019
&       1.4 \\
\civ\ $\lambda$ 1550 &  T4 & -90.0 & -45.0 &      0.663 &      0.024 &     0.023
&       1.1 \\
\civ\ $\lambda$ 1550 &  T5 & -40.0 &  15.0 &      0.368 &      0.036 &     0.031
&       1.2 \\
\civ\ $\lambda$ 1550 &  T6 &  20.0 &  70.0 &      0.829 &     -0.021 &     0.020
&     -1.1 \\
\civ\ $\lambda$ 1550 &  T7 &  95.0 &   155 &      0.881 &      0.004 &     0.018
&      0.2 \\
\civ\ $\lambda$ 1550 &  T8 &   165 &   200 &      0.971 &     -0.090 &     0.023
&     -3.9 \\
\civ\ $\lambda$ 1550 &  T9 &   205 &   250 &      0.978 &     -0.029 &     0.020
&     -1.4 \\
\hline
\label{coldensi}                         
\end{tabular}
\tablefoot{
\tablefoottext{a}{mean transmission in the COS spectrum.}
\tablefoottext{b}{mean fractional difference between COS and STIS troughs
normalized by the mean COS transmission.}
\tablefoottext{c}{mean fractional error in the difference between COS and STIS
troughs normalized by the mean COS transmission.}
\tablefoottext{d}{mean fractional difference between COS and STIS troughs
normalized by the error.}
}\end{table*}   

\subsection{column density determination}

  For each epoch, we determine the ionic column densities associated with the
nine
  components (T1--T9) shown in Figure~\ref{fig_cosstis} by modeling the
  residual intensity observed accross the absorption troughs. Assuming a single
homogeneous
  emission source $F_0$ whose spatial extension is normalized to 1, the
transmitted flux
  $F_i(v)$ for a line $i$ can be written as
    \begin{equation}
      F_i(v) = F_0(\lambda (v)) \int_{0}^{1} e^{-\tau_i(x,v)} dx 
    \end{equation}
  where $v$ is the radial velocity of the outflow and $\tau_i(x,v)$ is the
optical
  depth of the absorber accross the emission source. In this relation, we
implicitly
  reduced the number of spatial dimensions from two to one. This assumption,
whose validity is discussed in \citet{Arav05}, allows us to derive
  meaningful quantities from the fitting of residual intensity profiles. 
  We consider two common models for the absorber: the apparent optical depth
(AOD) model where the absorbing
  material is simply characterized by $\tau_i(x,v)=\tau_i(v)$ and fully covers
the emission source,
  and the partial-covering (PC) model in which the material with $\tau_i(v)$
only covers a fraction $C(v)$ of the emission source at a given velocity.
  Once computed over the width of the trough, the optical depth solution
$\tau_i(x,v)$
  is transformed into column density using the relation
  \begin{equation}
   N_{ion}(v) =  \frac{3.8 \times 10^{14}}{f_i \lambda_i} < \tau_i(v) > ~
(\mathrm{cm}^{-2} \: \mathrm{km}^{-1} \: \mathrm{s})
  \end{equation}
  where $f_i$, $\lambda_i$ and $< \tau_i(v) >$ are respectively the oscillator
strength, the
  rest wavelength and the average optical depth accross the emission source of
line $i$ (see \citealt{Edmonds11} for details). The main uncertainty in the
fitting procedure,
  and thus in the derived column density, comes from the assumption about the
spatial distribution of the absorbing material.

\begin{figure*}
  \includegraphics[angle=90,width=1.0\textwidth]{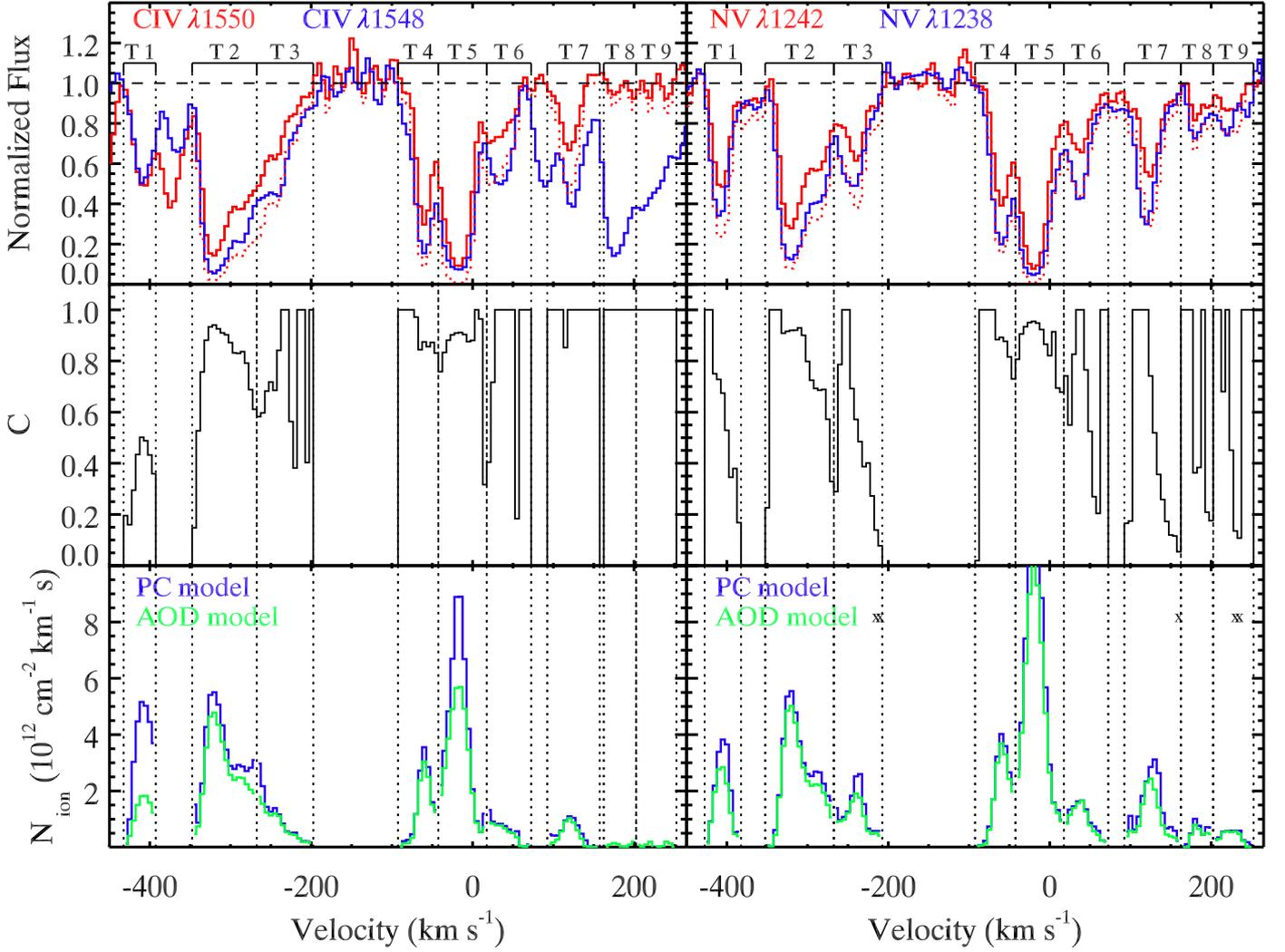}\\
 \caption{The upper panel presents the line profiles associated to \civ\ and
\nv\ in the COS spectrum. The dotted
  red line, only plotted in regions free of known blending in the weakest
transition (red line), represents the
  expected residual intensity in the stronger transition (blue line) assuming an
absorber totally covering the emission source (AOD model, $I_b = I_r^2$).
  The central panel shows the derived covering solution obtained for \civ\ and
\nv\ as a function of the radial velocity. A high covering is observed in most
components of the outflow, while trough T1 presents a lower covering in both
\civ\ and \nv. The low covering derived in troughs T6 and T7 of \nv\ is most
likely due to a shift in the wavelength solution.
  In the lower panel we plot the ionic column density solution derived for both
ions. The AOD model solution is computed on the red component of the doublet.
Velocity bins in which the PC solution has been corrected to account for the
shallowness of the trough are denoted with crosses (see text).}
 \label{lineprofplot} 
\end{figure*}

  The PC model is considered in order to account for the fact that in AGN, when
one observes at least two lines from the
  same ion, the apparent optical depth ratio $R_{ij}^{app}$ between the lines
$i$ and $j$ does not always
  follow the expected laboratory value $R_{ij}^{lab}=\lambda_i f_i / \lambda_j
f_j$. This observation can be explained if
  the absorber only partially covers the emission source (e.g.
\citealt{Hamann97b,Arav99a}). In the case of doublet
  lines like \civ, \nv, and \siIV, $R_{br}^{lab} \simeq 2$; i.e., the blue
transition of the doublet is twice as strong as the red one. Therefore, the
residual intensity in the blue line should lie between $I_b = I_r^2$ (AOD model)
and $I_b = I_r$ (fully saturated trough in the PC model).
  In the upper panels of Fig.~\ref{lineprofplot} we plot the COS \civ\ and \nv\
line profiles as well as the expected residual intensity
  for the strongest transition assuming the AOD scenario, highlighting the
allowed
  physical range of values for $I_b$ based on the observation of $I_r$. While
several kinematic components show a
  significant departure from the AOD prediction, suggesting a partial covering
of the emission source, none of
  the unblended components exhibits a strong saturation effect (i.e. $I_b \simeq
I_r$) with the exception of \civ\ in component T1. This allows us to determine
  accurate ionic column densities as well as quantitatively examine the
variations in the absorber between the
  STIS and the COS epochs. 
  
  In the lower panels of Fig.~\ref{lineprofplot} we display the $N_{ion}(v)$
solution derived for \civ\ and \nv\ for the COS epoch for both
  the AOD (computed on the weakest line of the doublet) and PC absorber models.
The PC solution can only be computed
  if one observes at least two unblended lines from the same ion. Given the
blending of the
  blue \civ\ line in trough T7 and the non-detection of \civ\ in troughs T8 and
T9 we do not report column for these troughs using the PC model, but only
provide a lower limit based on the AOD model. Trough T1
  is strongly saturated in \civ\ as revealed by the perfect match of the shape
of the blue and red component line profiles.
  This only allows us to place a conservative lower limit on the column density
by assuming an optical depth of at least $\tau = 4$
  across the line profile of the red component. Looking at the \siIV\ line
profile, only detected
  in trough T5, reveals an optically thin absorption line with a covering of
unity accross the trough.
  We corrected the PC solution in several velocity bins by using the mean PC
solution derived in the adjacent pixels in
  order to account for the increased sensitivity of the PC solution to the noise
when modeling the shallower parts of the troughs.
  These points are marked with crosses in Fig.~\ref{lineprofplot}.  

  We list the integrated values of the computed column densities across the nine
independent components using both absorber models
  in Table~\ref{line_data}. Except for \civ\ in component T1, the integrated
column densities obtained using the two absorber models
  are generally in agreement (with differences $\ltsim 30\%$) for both STIS and
COS data lending further support to the non-saturation
  of the components. A higher discrepancy is generally observed for shallow
troughs in the STIS spectrum and is explained
  by the lower S/N in that dataset (cf. the \siIV\ measurement).
  In Table~\ref{line_data}, we also provide the differences in column densities
determined between the COS and STIS epochs as well as the
  fractional differences in column densities normalized to the COS measurement
for both absorber models.
  One can see that the fractional differences observed are in general agreement
between both absorber models, as already suggested by the 
  small difference in computed column density typical of non-saturated troughs.


\begin{table*}[!tbp]
\caption{computed column densities} 
\centering                
\begin{tabular}{ll|llcl|llcl} 
\hline\hline
 & & \multicolumn{4}{|c}{AOD} & \multicolumn{4}{|c}{PC} \\
Trough  & Ion & C\tablefootmark{a} & S\tablefootmark{b} & C-S & (C-S)/C & C & S
& C-S & (C-S)/C \\
\hline
\vspace{0.6mm}
T1 & N(\civ)  & $>$ 47.1         & $>$ 39.0               &  $\cdots$ & 
$\cdots$                  &  $>$ 135             &  $>$ 101                  &  
$\cdots$   &   $\cdots$  \\
\vspace{0.6mm}
T1 & N(\nv)   &     64.8$\pm0.8$ &     48.6$\pm5.0$       &   16.2    &   
0.25$\pm0.08$           &  86.4$\pm1.5$        &     70.7$^{+18}_{-8.7}$   &  
15.8   &    0.18$^{+0.21}_{-0.10}$ \\
\vspace{0.6mm}
T2 & N(\civ)  &     227$\pm1$    &     252$\pm8$         &  -25      &  
-0.11$\pm0.03$           &  264$\pm1$           &      288$^{+54}_{-13}$    & 
-24.0   &   -0.09$^{+0.20}_{-0.05}$ \\
\vspace{0.6mm}
T2 & N(\nv)   &     201$\pm1$    &     215$\pm10$        &  -14      &  
-0.07$\pm0.05$           &  225$\pm1$           &      254$^{+480}_{-13}$   & 
-28.4   &   -0.13$^{+2.1}_{-0.06}$ \\
\vspace{0.6mm}
T3 & N(\civ)  &     56.0$\pm0.4$ &     61.4$\pm3.4$       &   -5.4    &  
-0.10$\pm0.06$          &  68.9$\pm0.7$        &     67.9$\pm3.9$         &  
1.0    &   0.01$\pm0.06$ \\
\vspace{0.6mm}
T3 & N(\nv)   &     62.2$\pm0.8$ &     64.6$\pm5.6$       &   -2.5    &  
-0.04$\pm0.09$           &  78.0$\pm1.2$        &     84.5$^{+670}_{-8.9}$  &  
-6.5   &   -0.08$^{+8.6}_{-0.16}$ \\
\vspace{0.6mm}
T4 & N(\civ)  &     64.1$\pm0.5$ &     71.4$\pm3.4$       &   -7.3    &  
-0.11$\pm0.05$           &  72.7$\pm0.5$        &     76.5$\pm3.1$          &  
-3.8   &   -0.05$\pm0.04$ \\
\vspace{0.6mm}
T4 & N(\nv)   &     84.7$\pm0.9$ &     78.5$\pm5.4$       &     6.1   &  
0.07$\pm0.06$           &  92.7$^{+38}_{-0.8}$ &     91.2$^{+8.4}_{-6.4}$  &  
1.5    &   0.02$^{+0.42}_{-0.07}$ \\
\vspace{0.6mm}
T5 & N(\civ)  &     197$\pm1$    &     211$\pm10$        &  -13      &  
-0.07$\pm0.05$           &  264$\pm4$          &      266$^{+69}_{-13}$    &  
-2.0   &   -0.01$^{+0.26}_{-0.05}$ \\
\vspace{0.6mm}
T5 & N(\nv)   &     299$\pm2$    &     272$^{+22}_{-14}$  &     28    &  
0.09$^{+0.08}_{-0.05}$   &  356$\pm5$           &      346$^{+2500}_{-34}$  &  
10.2   &   0.03$^{+6.9}_{-0.10}$ \\
\vspace{0.6mm}
T5 & N(\siIV) &    2.7$\pm0.2$   &    4.6$^{+1.4}_{-1.0}$ &   -2.0    &  
-0.74$^{+0.54}_{-0.38}$  &  2.8$^{+21}_{-0.1}$  &      9.5$^{+43}_{-1.9}$   &  
-6.7   &   -2.5$^{+31}_{-0.7}$ \\
\vspace{0.6mm}
T6 & N(\civ)  &     28.2$\pm0.4$ &     28.9$\pm2.7$       &   -0.7    &  
-0.03$\pm0.10$           &  33.7$\pm0.5$        &     35.3$^{+5.6}_{-3.4}$  &  
-1.6   &   -0.05$^{+0.17}_{-0.10}$ \\
\vspace{0.6mm}
T6 & N(\nv)   &     50.6$\pm0.8$ &     48.4$\pm4.9$       &     2.3   &  
0.04$\pm0.10$            &  54.8$\pm0.7$        &     53.6$\pm3.8$          &   
1.2   &   0.02$\pm0.07$ \\
\vspace{0.6mm}
T7 & N(\civ)  &     24.9$\pm0.4$ &     27.7$\pm2.8$       &   -2.8    &  
-0.11$\pm0.12$           & $\cdots$             & $\cdots$                  &
$\cdots$ & $\cdots$  \\
\vspace{0.6mm}
T7 & N(\nv)   &     64.4$\pm0.5$ &     66.7$\pm4.5$       &     -2.3  &   
-0.04$\pm0.07$          &  96.8$\pm2.3$        &     69.0$^{+360}_{-5.9}$  &   
27.9  &    0.29$^{+3.8}_{-0.06}$ \\
\vspace{0.6mm}
T8 & N(\civ)  & $<$ 3.7          & $<$ 5.2                & $\cdots$  & 
$\cdots$                  &  $\cdots$            &  $\cdots$                 &
$\cdots$ &  $\cdots$ \\
\vspace{0.6mm}
T8 & N(\nv)   &     16.8$\pm0.6$ &     16.6$\pm3.6$       &    0.1    & 
0.01$\pm0.22$             &  21.3$\pm0.8$        &     24.2$^{+290}_{-7.9}$  &  
-2.9   &   -0.14$^{+14}_{-0.37}$ \\
\vspace{0.6mm}
T9 & N(\civ)  & $<$ 4.2          & $<$ 4.9                & $\cdots$  & 
$\cdots$                  & $\cdots$             & $\cdots$                  &
$\cdots$ & $\cdots$ \\
\vspace{0.6mm}
T9 & N(\nv)   &     17.4$\pm0.7$ &     21.9$\pm4.2$       &   -4.5    &  
-0.26$\pm0.25$           &  20.5$\pm0.5$        &     34.0$^{+150}_{-9.4}$  & 
-13.5   &   -0.66$^{+32}_{-0.46}$ \\
\hline
\label{line_data}                         
\end{tabular}
\tablefoot{
\tablefoottext{a}{$N_{ion}$ in units of $10^{12}$ cm$^{-2}$ measured from the
2009 COS observations. All errors are statistical only.}
\tablefoottext{b}{$N_{ion}$ in units of $10^{12}$ cm$^{-2}$ measured from the
2001 STIS observations.}
}\end{table*}   

\section{Photoionization Solutions for the Different Outflow Components of Mrk
509}

Our distance determinations rely on knowledge of the ionization parameter, which
we find by solving the photoionization and thermal equilibrium equations
self-consistently using version c08.00 of the spectral synthesis code {\sc
Cloudy} (last described by \citealt{Ferland98}). We use the spectral energy
distribution (SED) described in Paper I and assume a plane-parallel geometry, a
constant hydrogen number density, and solar abundances as given in {\sc Cloudy}.
These abundances differ from those of \citet{Lodders09} (see
Table~\ref{tab:abundances}) used in Paper VIII, but the differences do not
significantly affect our results. Grids of models are generated where the total
hydrogen column density ($N_H$) and the ionization parameter ($U_H$) are varied
in 0.1 dex steps (similar to the approach of \citealt{Arav01,Edmonds11}) for a
total of $\sim$4500 grid points covering a parameter space with 15 $\le$ log
$N_H$ $\le$ 24.5 and -5 $\le$ log $U_H$ $\le$ 2. Intermediate values are
estimated by a log interpolation. At each point of the grid, we tabulate the
predicted column densities ($N_{ion}$) of all relevant ions and compare them
with the measured column densities (see Table~\ref{line_data}). Our solutions
are based only on \civ\ and \nv\ (except for trough T5 discussed below). These
lines cross at a single point in the $N_H,U_H$ plane yielding a unique solution.
The results for both COS and STIS data are given in Table~\ref{table:models}.
For most components, the differences in log~$N_H$ and log~$U_H$ between the four
determinations (AOD and PC for both COS and STIS) are around 0.1--0.2 dex, and
therefore do not affect our distance limits. The exception is component T1 where
the AOD and PC determinations are significantly different due to the saturation
of \civ. We obtain a photoionization solution for T1 by determining the upper
limit on $N$(\siIV) along with the lower limit on $N$(\civ) and the measurement
of $N$(\nv). In the last two columns of Table~\ref{table:models}, we
give the fractional difference in column densities expected if the number
density were high enough for the absorber to be in photoionization equilibrium
at the time of the COS observations. Comparison with the sixth column of
Table~\ref{line_data} reveals that the absorber is out of equilibrium.

\begin{table}[!tbp]
\caption{chemical abundances} 
\centering                
\begin{tabular}{ccc} 
\hline\hline
Element & \textsc{Cloudy}\tablefootmark{a} & \textsc{Lodders
2009}\tablefootmark{b} \\
\hline
 He & -1.00 & -1.07  \\
 C  & -3.61 & -3.61  \\
 N  & -4.07 & -4.14  \\
 O  & -3.31 & -3.27  \\
 Ne & -4.00 & -3.95  \\
 Mg & -4.46 & -4.46  \\
 Si & -4.46 & -4.47  \\
 S  & -4.74 & -4.84  \\
 Fe & -4.55 & -4.54  \\
\hline
\label{tab:abundances}                         
\end{tabular}
\tablefoot{Abundances are given in log relative to hydrogen with
log[A(H)]=0.00.
\tablefoottext{a}{Abundances as given in \textsc{Cloudy} used in this paper.}
\tablefoottext{b}{\citet{Lodders09} abundances used in Paper VIII.}
}\end{table}   

In trough T5, \siiv\ is detected in addition to \civ\ and \nv. \citet{Kraemer03}
concluded that two ionization parameters are needed to match the observational
constraints for this trough (their component 4) under the assumption of solar
abundances. Two models are presented in Table~\ref{table:models} for trough T5,
one for each ionization component. The high ionization model fitting \civ\ and
\nv\ underpredicts \siiv\ by a factor of $\sim 10$ which is ameliorated by the
addition of a lower ionization component fitting \siiv. Summation of the
predicted column densities for the two ionization components results in an
overprediction of \civ\ by a factor of 2. The solution is improved by increasing
$U_H$ and $N_H$ of the high ionization component resulting in a band of
solutions with log $U_H \gtsim -1.0$. These models predict all of the \siiv\ and
\civ\ come from the low ionization component, while \nv\ comes from the high
ionization component. Our results differ from those of \citet{Kraemer03},
especially in the low ionization component, where they find about 10 times
larger $N_H$ than we do. They find such a large value by assuming a low covering
of the emission source by \siiv. With higher S/N COS data, however, we find a
covering near unity. We assume the high and low ionization components are at the
same location, an assumption supported by the kinematic correspondence of all
three troughs, and use the low ionization component solution to provide a lower
limit on the distance.
 
It is also possible to find a single ionization parameter solution for component
T5 if the assumption of solar abundances is relaxed. We find that increasing the
abundances of nitrogen and silicon relative to carbon by a factor of 2 results
in a model that accurately predicts the column densities of \civ, \nv, and
\siiv, with $\log U_H = -1.5$, and $\log N_H = 18.7$, values close to the low
ionization component discussed above. However, \citet[Paper VII]{Steenbrugge11}
used \textit{XMM-Newton} and \textit{Chandra} data to show that the abundances
for C, N, and Si are consistent with the proto-solar abundances determined by
\citet{Lodders09}, and the ratio of nitrogen to carbon abundances is less than
30\% higher than the solar ratio.

For each kinematic component (except T5), we find a satisfactory fit to the data
with a single ionization component. This differs from the X-ray analysis in
\citet[hereafter Paper III]{Detmers11} where some ions are formed by multiple
ionization components. However, since the velocities are not resolved in the
X-ray spectra, this does not necessarily imply disagreement between the X-ray
and UV analysis. A comprehensive comparison of the UV and X-ray data is deferred
to a future paper (Ebrero et al. 2012, in preparation).

\begin{table*}[!tbp]
\caption{photoionization models} 
\centering                
\begin{tabular}{lcccccccccc} 
\hline\hline
 & \multicolumn{4}{c}{COS} & \multicolumn{4}{c}{STIS} &
\multicolumn{2}{c}{$\frac{\Delta N_{ion}}{N_{ion} \textrm{(STIS)}}$\tablefootmark{a}} \\
 & \multicolumn{2}{c}{AOD} & \multicolumn{2}{c}{PC} & \multicolumn{2}{c}{AOD} &
\multicolumn{2}{c}{PC} &  &  \\
Trough & log $U_H$ & log $N_H$ & log $U_H$ & log $N_H$ & log $U_H$ & log $N_H$ &
log $U_H$ & log $N_H$ & \textsc{Civ} & \textsc{Nv} \\
 & & (cm$^{-2}$) & & (cm$^{-2}$) & & (cm$^{-2}$) & & (cm$^{-2}$) & & \\
\hline
 T1 & -1.1 & 18.5 & -1.5 & 18.5 & -1.2 & 18.3 & -1.5 & 18.4 & -0.51 & -0.17 \\
 T2 & -1.4 & 18.8 & -1.4 & 18.9 & -1.4 & 18.9 & -1.3 & 19.0 & -0.55 & -0.28 \\
 T3 & -1.2 & 18.4 & -1.2 & 18.6 & -1.3 & 18.4 & -1.1 & 18.7 & -0.57 & -0.35 \\
 T4 & -1.1 & 18.6 & -1.1 & 18.6 & -1.2 & 18.5 & -1.2 & 18.6 & -0.60 & -0.41 \\
 T5(high)\tablefootmark{b} & -1.0 & 19.2 & -1.1 & 19.2 & -1.1 & 19.1 & -1.1 &
19.2 & -0.62 & -0.48 \\
 T5(low) & -1.6 & 18.6 & -1.5 & 18.2 & -1.8 & 18.4 & -1.5 & 18.2 & -0.41 & +0.26
\\
 T6 & -0.9 & 18.6 & -0.9 & 18.6 & -0.9 & 18.5 & -0.9 & 18.8 & -0.68 & -0.56 \\
 T7\tablefootmark{c} & -0.5 & 19.2 & $\cdots$ & $\cdots$ & -0.8 & 18.8 &
$\cdots$ & $\cdots$ & -0.72 & -0.59 \\
\hline
Total $N_H$ & $\cdots$ & 19.8 & $\cdots$ & 19.6 & $\cdots$ & 19.6 & $\cdots$ &
19.7 &  &  \\
\hline
\label{table:models}                         
\end{tabular}
\tablefoot{Troughs T8 and T9 have heavy blending in the blue component of
\civ, and the red component of \civ\ is very weak. We therefore do not include
ionization analysis of these troughs.
\tablefoottext{a}{Fractional changes in column density expected if the
number density were high enough for the absorber to be in equilibrium at the
time of the COS observations (see section 3)}
\tablefoottext{b}{Lower limits. Summation of low and high ionization components
for trough T5 overpredicts \civ\ by a factor of 2, which is ameliorated by
increasing $U_H$ of the high ionization component. Since all of the \civ\ comes
from the low ionization component, we use the lower ionization parameter to
compute lower limits on the distance for this trough (see Section 3).}
\tablefoottext{c}{Heavy blending in the blue component of \civ\ precludes
partial covering measurements for trough T7.}
}\end{table*}   

\begin{table*}[!t]
\caption{initial values} 
\centering                
\begin{tabular}{lcccccccc} 
\hline\hline
Trough & $\alpha$(\textsc{Ciii}) & $\alpha$(\textsc{Civ}) & log
$N$(\textsc{Civ}) & log $N$(\textsc{Cv}) & $\alpha$(\textsc{Niv}) &
$\alpha$(\textsc{Nv}) & log $N$(\textsc{Nv}) & log $N$(\textsc{Nvi}) \\
 & \multicolumn{2}{c}{($10^{-12}$ cm$^3$ s$^{-1}$)} & 
\multicolumn{2}{c}{(cm$^{-2}$)} & \multicolumn{2}{c}{($10^{-12}$ cm$^3$
s$^{-1}$)} & \multicolumn{2}{c}{(cm$^{-2}$)} \\
\hline

 T1 & 23.9 & 5.86 & 14.1 & 14.8 & 28.0 & 10.8 & 14.0 & 13.9  \\
 T2 & 24.5 & 5.66 & 14.3 & 15.1 & 28.2 & 10.3 & 14.3 & 14.3  \\
 T3 & 26.4 & 5.26 & 13.7 & 14.7 & 28.9 & 9.40 & 13.8 & 14.1  \\
 T4 & 27.6 & 5.07 & 13.8 & 14.9 & 29.5 & 9.00 & 13.9 & 14.3  \\
 T5(low) & 23.3 & 6.09 & 14.1 & 14.8 & 27.9 & 11.3 & 14.1 & 13.9  \\
 T6 & 30.8 & 4.67 & 13.5 & 14.8 & 31.4 & 8.23 & 13.7 & 14.4  \\
 T7 & 41.1 & 3.73 & 13.4 & 15.2 & 39.7 & 6.53 & 13.9 & 15.0  \\

\hline
\label{table:iv}                         
\end{tabular}
\tablefoot{Using $N_H$ and $U_H$ derived from the 2009 COS data (see
text) using the AOD method except for component T1 where we use values from the
PC method due to the saturation of \civ.
}\end{table*}   

\section{Time-dependent Ionization Equations}

The ionization parameter
\begin{equation}
\label{eqionparam}
U_H \equiv {Q_H\over {4\pi R^2 \vy{n}{H} c}}
\end{equation}
(where $Q_H$ is the rate of hydrogen ionizing photons emitted by the central
source, $c$ is the speed of light, $R$ is the distance to the absorber from the
central source, and $\vy{n}{H}$ is the total hydrogen number density)
characterizes a plasma in photoionization equilibrium. When the ionizing flux
varies, the ionization state of the gas will change in response if the timescale
for flux variations is an appreciable fraction of the recombination timescale
for the gas. The latter depends on the electron number density ($n_e$), which is
$\simeq 1.2 \vy{n}{H}$ in highly ionized plasma. Gases of high density will
respond faster than gases of low density due to a higher collision rate between
free electrons and ions (e.g., \citealt{Krolik95,Nicastro99}; Paper VIII). If
the gas has not had time to reach ionization equilibrium, determination of $U_H$
by line ratios suffers from uncertainties since it is inappropriate to use the
assumption of photoionization equilibrium. As we show in the appendix, in that
case, the ionization state of the gas will be more accurately derived by using
the average $Q_H$ over a timescale roughly equal to the recombination timescale
of the ion in question. Tracking changes in column density of a given ion
between different epochs along with flux monitoring can lead to estimates of
$\vy{n}{H}$ and thereby, the distance $R$ (e.g., \citealt{Gabel05b}) assuming
that changes in the hydrogen number density between epochs is negligible. 

The abundance of a given element in ionization stage $i$ is given by
\begin{equation}
 \frac{dn_i}{dt}=-\vy{n}{i}(I_i+R_{i-1})+n_{i-1} I_{i-1} + n_{i+1} R_i,
\label{eq:ionfrac}
\end{equation}
as a function of the ionization rate per particle, $I_i$, and the recombination
rate per particle from ionization stage $i+1$ to $i$, $R_i$. We have neglected
Auger effects, collisional ionization, and charge transfer (e.g.,
\citealt{Krolik95}). If the gas at distance $r$ from an ionizing source of
monochromatic luminosity $L_\nu$ is optically thin, as in Mrk 509, the
ionization rate per particle is given by
\begin{equation}
I_i=\int_{\nu_i}^\infty\frac{(L_\nu/h\nu)\sigma_\nu }{4\pi r^2}d\nu,
\label{eq:irate}
\end{equation}
where $h$ is Planck's constant and $\sigma_\nu$ is the cross-section for
ionization by photons of energy $h\nu$. The recombination rate per particle is
given by
\begin{equation}
R_i=\alpha_i(T)n_e.
\label{eq:rrate}
\end{equation}
The recombination coefficient $\alpha$ depends on the electron temperature $T$
and scales roughly as $T^{-1/2}$ \citep{AGN^3}.

Equation~\ref{eq:ionfrac} forms a set of $n+1$ coupled differential equations
for an element with $n$ electrons and $n+1$ ions. In the steady state, these
reduce to $n$ equations of the form
\begin{equation}
\frac{n_{i+1}}{n_i}=\frac{I_i}{R_i}.
\label{eq:steadystate}
\end{equation}
Closure of the steady state set of equations is given by $\sum \: n_i =
n_{tot}$, where $n_{tot}$ is the total number density of the element in
question. Under these assumptions the level of ionization of the gas in
photoionization equilibrium may be characterized by $I_i/R_i$, which is
proportional to the ratio of ionizing flux to $n_e$ and leads to the definition
of ionization parameter given in Equation~\ref{eqionparam}.

Simple scaling of Equation~\ref{eq:ionfrac} leads to a characteristic timescale.
Suppose an absorber in photoionization equilibrium experiences a sudden change
in the incident ionizing flux such that $I_i(t>0) = (1+f)I_i(t=0)$, where $-1
\le f \le \infty$. Then taking the ratio $dn_i / dt \rightarrow n_i / t$ leads
to the timescale for change in the ionic fraction:
\begin{equation}
 t^* = \left[ -f \alpha_i n_e \left( \frac{n_{i+1}}{n_i}-
\frac{\alpha_{i-1}}{\alpha_i} \right) \right]^{-1}.
\label{eq:trec}
\end{equation}
Note that the timescale defined here equals the recombination timescale of
\citet{Krolik95} when $f=-1$, i.e., the ionizing flux drops to zero (see also
\citealt{Nicastro99,Bottorff00,Steenbrugge09}). Including the ionizing flux in
$t^*$ gives more accurate timescales in cases where the ionizing flux either
changes by small amounts ($|f| \ll 1$) or increases by a large amount. The
recombination coefficients are obtained for each of our photoionization models
using the \textsc{Cloudy} command ``\texttt{punch recombination coefficients}''.
The initial values needed to compute recombination times for \civ\ and \nv\ for
troughs T1 through T7 are given in Table~\ref{table:iv}. We compute $t^*$ for
reference (see Table~\ref{table:rectimes}), but the distance determinations
discussed in Section~\ref{sec:simulations} are from explicit solutions of the
time-dependent photoionization equations using simulated lightcurves, not
timescale arguments. We use the values derived from the 2009 COS data since the
higher S/N allows for better constraints than the 2001 STIS data, and the
photoionization solutions are similar for both data sets (see
Table~\ref{table:models}). Results for troughs T8 and T9 are not given due to
heavy blending in the blue component of \civ\ and very weak lines in the red
component precluding reliable photoionization solutions. 

\begin{table}[!tbp]
\caption{timescales per electron number density} 
\centering                
\begin{tabular}{lcc} 
\hline\hline
Trough & $-f n_e t$(\textsc{Civ}) & $-f n_e t$(\textsc{Nv}) \\
 & ($10^{10}$ cm$^{-3}$ s) & ($10^{10}$ cm$^{-3}$ s) \\
\hline

 T1 & 18.3 & -5.15  \\
 T2 & 8.92 & -5.59    \\
 T3 & 3.82 & -9.86    \\
 T4 & 2.76 & -14.5    \\
 T5(low) & 13.8 & -4.81   \\
 T6 & 1.60 & 10.2 \\
 T7 & 0.51 & 2.35    \\

\hline
\label{table:rectimes}                         
\end{tabular}
\tablefoot{The numbers in this table are the product of $t^*$ from
Equation~\ref{eq:trec} and $-f n_e$. To get a timescale for a given $n_e$ and flux
change $f$, we divide the number in the table by $-f n_e$. For example, given
$n_e = 10^3$~cm$^{-3}$ (typical for the upper limits derived in Section 5) and
$f=0.72$ (the flux at the COS epoch minus that at the STIS epoch),
$t^*$(\civ)$\sim 1.7$~yr for component T3. Note that the numbers in this table are
positive or negative depending on whether the change in ionic fraction is
anti-correlated or correlated with a drop in ionizing flux, respectively (see
also \citealt{Bottorff00}).
}\end{table}   

It is common to use the recombination timescale ($-f t^*$;
  e.g.  \citealt{Krolik95,Bottorff00,Netzer08}) when determining
  limits on the number density of an AGN outflow. For large increases
  in flux, the ionization timescale ($I_i^{-1}$) has been invoked
  (e.g. \citealt{Dunn10b}).  Use of our refined timescale
  (Equation~\ref{eq:trec}) allows us to treat both increases and
  decreases in flux for any ion and account for finite flux changes in
  a natural way.

For the Mrk 509 UV data, there are two physically motivated
  timescales we can use in Equation~(\ref{eq:trec}): 1) assuming an
  instantaneous increase in flux just after the STIS epoch that stays
  constant through the COS epoch ($t^* = 8$ years and $f=0.72$ for
  this case); and 2) assuming a constant flux at the STIS epoch level
  until the 100 days monitoring prior to the COS observations followed
  by an instantaneous flux increase to the COS flux level thereafter,
  ($t^* = 100$ days and $f=0.72$ for this case). Using the appropriate
  ionization equilibrium for each component, we derive
  upper limits on the number density for each of these cases (see columns 3 and 5 of Table~\ref{tab:timescale_densities}). Due to the difference in timescales, the first case yields
  upper limits that are a factor of 30 smaller than the second
  case. We then use Equation~\ref{eqionparam} to derive the associated
  lower limits on the distance to the absorbers from the central
  source (see columns 4 and 6 of Table~\ref{tab:timescale_densities})

However, there are several limitations when using timescale arguments
in order to infer the number density (or limits thereof) of the
absorber.  First, timescale analysis
implicitly relies on the physically implausible lightcurves discussed above.
As we show in Section 5, a more physically motivated approach is to use 
lightcurve simulations that are anchored in our knowledge of the
power spectrum behavior of observed AGN lightcurves.

Second, timescale analysis does not take into account the quality of
measurement.  This is especially important for cases where no changes
in column density are observed.  We expect that tighter
error bars on no-change measurements would yield smaller upper limits
on the absorber's $n_e$.  To correct the timescale inferred $n_e$  values
for this effect we use the following approach. For  the simple
lightcurve associated with the 100 days timescale, we
numerically solve equation set \ref{eq:ionfrac}, while requiring that changes in
ionic column densities are less than the 1-$\sigma$ errors from
Table~2. The resulting limits on $n_e$ and $R$ are given in columns
7 and 8 of Table~\ref{tab:timescale_densities}, designated $n_e3$ and $R3$. We note that we are able to put a range on the density for T1 and T2 due to the observed change in column density for these components.

Third, use of Equation~\ref{eq:trec} can lead to
  problems when used for ions near their maximum concentration and
  should be avoided in these cases. As discussed in Paper VIII, ions
  near their maximum concentration are relatively insensitive to
  ionizing flux changes. In these cases, using Equation~\ref{eq:trec}
  can result in a large overestimation of the electron number
  density. For example, for an electron number density of
  $1900$~cm$^{-3}$, the \civ\ timescale for trough T2 is $\sim 5$
  times larger than the e-folding time determined by solving
  Equation~\ref{eq:ionfrac} numerically.

\begin{table}[!tbp]
\caption{density and distance limits from timescale calculations} 
\centering                
\begin{tabular}{c@{\,}c@{\,}c@{\,}c@{\,}cccc} 
\hline\hline
Trough & $v$ & log $n_e1$ & $R1$ & log $n_e2$ & $R2$ & log $n_e3$ & $R3$ \\
 & (km s$^{-1}$) & (cm$^{-3}$) & (pc) & (cm$^{-3}$) & (pc) & (cm$^{-3}$) & (pc)\\
\hline
 T1 & -405 & $<$2.4 & $>$460  & $<$3.9 & $>$80  & 3.0--3.7 & 100--230  \\
 T2 & -310 & $<$2.4 & $>$400 & $<$3.9 & $>$70  & 2.9--3.1 & 180--230  \\
 T3 & -240 & $<$2.3 & $>$350 & $<$3.7 & $>$70  & $<$2.7 & $>$220  \\
 T4 & -70  & $<$2.1 & $>$400 & $<$3.6 & $>$70  & $<$2.6 & $>$230  \\
 T5 & -15  & $<$2.4 & $>$490 & $<$3.8 & $>$100 & $<$3.6 & $>$120  \\
 T6 & +45  & $<$1.9 & $>$430 & $<$3.4 & $>$80  & $<$1.8 & $>$480  \\
 T7 & +125 & $<$1.4 & $>$480 & $<$2.9 & $>$90  & $<$1.9 & $>$270  \\
\hline
\label{tab:timescale_densities}                         
\end{tabular}
\tablefoot{
Timescales are derived from \civ\ for all troughs except T1, for which we use \nv. While troughs T1 and T2 show change, the limits on number density derived from timescale arguments are upper limits since photoionization models imply the absorbers are not in equilibrium.
}\end{table}

\begin{figure}[!t]
 \includegraphics[angle=90,width=0.7\hsize]{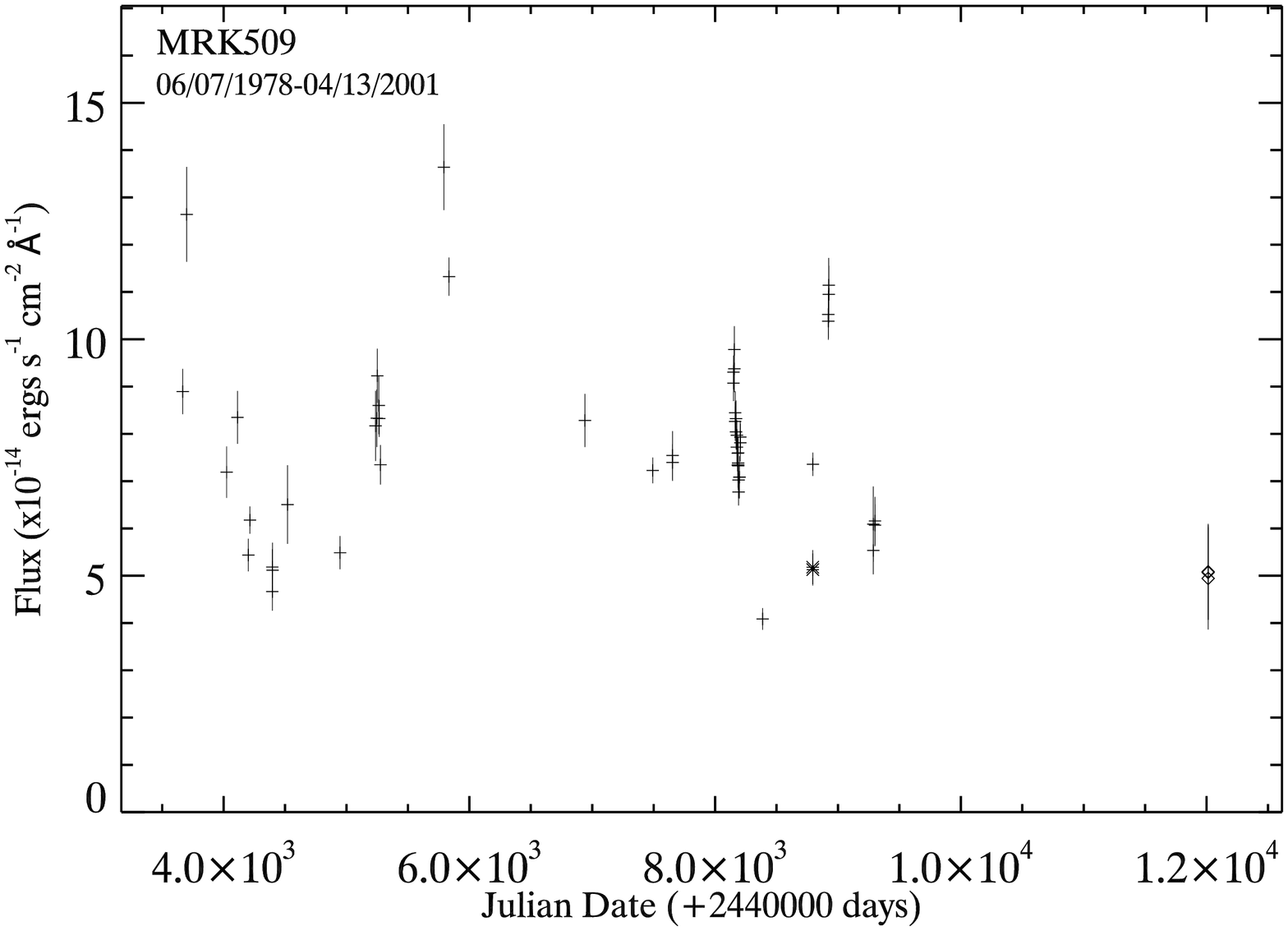}\\
  \includegraphics[angle=90,width=0.7\hsize]{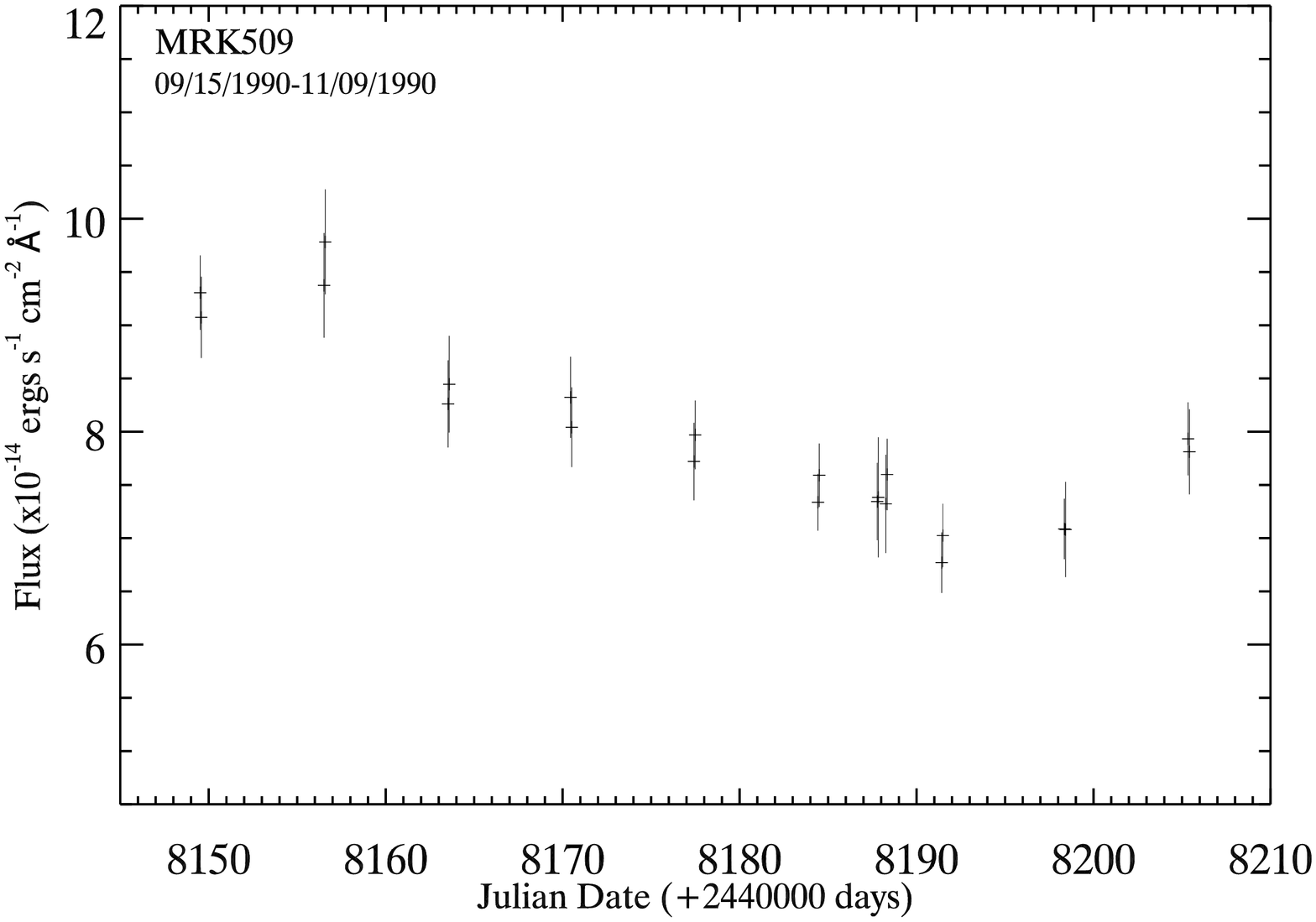}\\
 \includegraphics[angle=90,width=0.7\hsize]{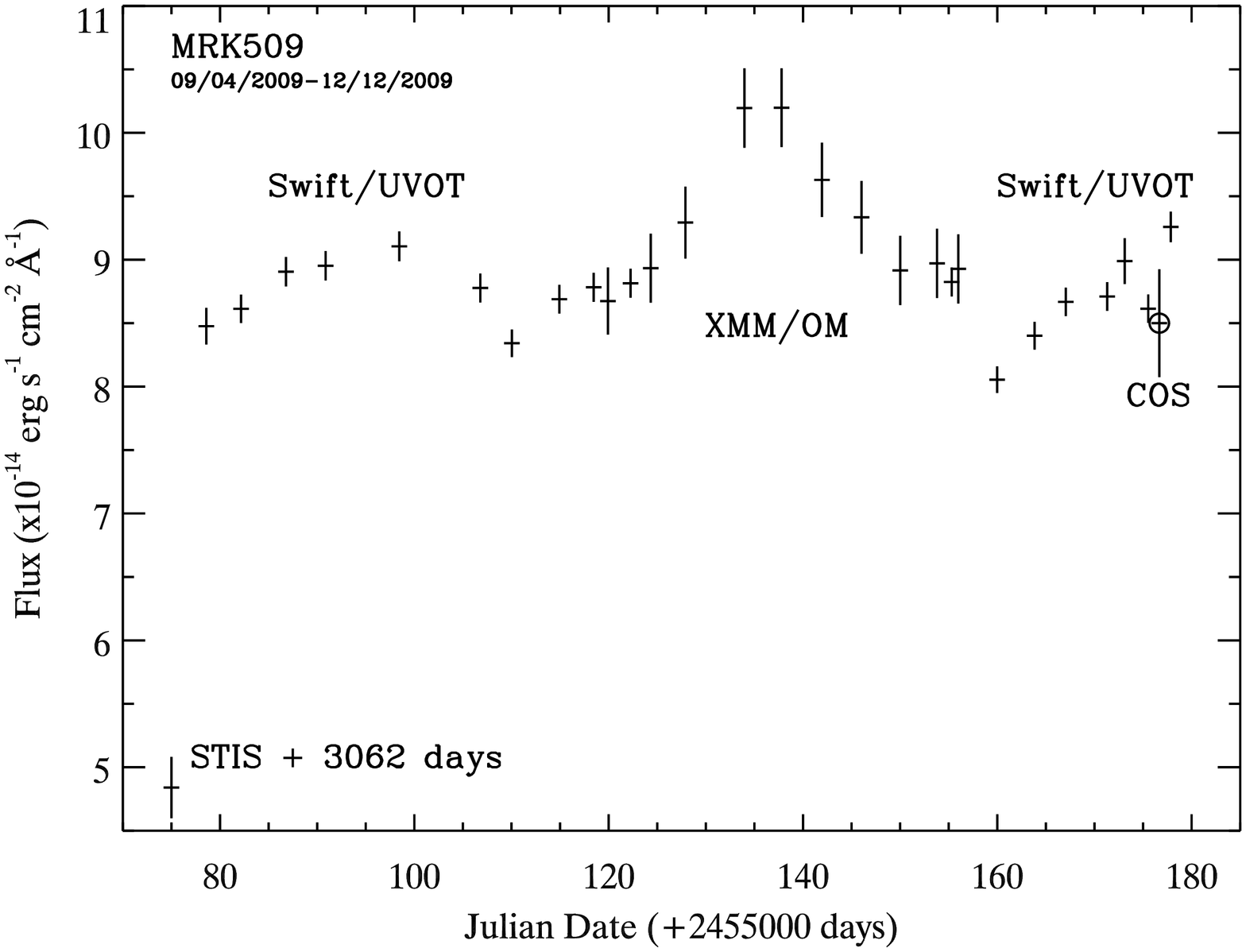}\\
 \caption{MRK 509 UV lightcurve monitoring at 1401$\AA$. The top
panel shows all the UV flux monitoring observations of Mrk 509 prior to our
campaign. Even in this sparse sampling, it is clear that the UV lightcurve of
Mrk 509 changes by at least a factor of 3.5. The center panel gives a blow-up of
the only intensive short-term monitoring (September-November
1990) before our campaign. The characteristics of this intensively monitored lightcurve are similar
to the one we measured during our 2009 campaign shown in the bottom panel, where the COS observation is marked with a circle.}
 \label{fig:lc}
\end{figure}

\section{Monte Carlo Simulations of Absorption Trough Changes}
\label{sec:simulations}

 As mentioned above, timescale analysis implicitly relies on
  physically implausible lightcurves. This could be justified
  if these lightcurves resulted in ``conservative'' or ``robust''
  limits on $n_e$. However, designating the 100 days timescale
  lightcurve (see section 4) as case A, we give examples of two
 cases that give larger upper limits on $n_e$: \\
Case B: The UV flux dropped to a very low state (say, 1\% of the STIS flux level) shortly after the STIS epoch and instantaneously jumped to the COS level 100 days prior to that epoch (the time period for which we have monitoring). In this case the resulting $n_e$ will be larger than in Case A. For example, solving Equation~\ref{eq:ionfrac} with this lightcurve results in $n_e$ that is larger by a factor of 5 for kinematic component T2. \\
Case C. The flux level before the STIS
measurement was similar to the one measured at the COS epoch and dropped suddenly just before the STIS observation, returning to the COS level shortly after the STIS epoch. In this case, we
lose any distance information since no change in column density is
expected and thus the electron number density could be arbitrarily
high.

 However, historical UV monitoring data of both Mrk 509
 (see figure 3) and other nearby AGN clearly show that all 3 cases
 discussed above are highly unlikely. We therefore use a different approach to assess the limits on
  $n_e$ from the available data and the well studied power spectrum
  behavior of AGN lightcurves.  Using this information, we are able
  to produce representative simulated lightcurves that allow us to
  derive the (physical) statistical constraints on the upper limits
  for the number density of the outflow and therefore lower limits on
  the distance.  This method also offers inherent improvements on
  traditional timescale analysis by accounting for the physical
  behavior AGN lightcurves and the quality of measurement in cases
  where no changes are observed in the absorption troughs (see second
  limitation of timescale analysis in section 4). We will show in
  section 6 that the two simple applications of the timescale
  described in the previous section are similar to and also bracket
  the statistical limits we obtain in this section.

There was no monitoring of the lightcurve of Mrk 509 between the STIS
observation in 2001 and our 2009 multiwavelength campaign. However, we can use
the prior history of UV and optical
monitoring of Mrk 509 to establish the expected character of any variations
that might have occurred.
In general, the optical and UV continua show variations that are well
characterized by a power-law power density spectrum $P(f) \propto f^{-\alpha}$,
with spectral indices in the range of 1 to 2.5
\citep{White94,Peterson98,Collier01,Horne04}.
\citet{Collier01} analyzed ground-based optical monitoring data for
Mrk 509 as part of a study to characterize the optical and UV continuum
variations of AGN. For 1908 days of monitoring at 10 to 100 day
intervals, they established that the power density spectrum of Mrk 509 has
a spectral index of $\alpha = 2.06 \pm 0.14$.
To see what such variations over the 8 years between the STIS and COS
observations might imply for changes in the UV-absorbing gas, we
perform a Monte Carlo simulation to generate a set of 1000 light curves using
the variability characteristics of Mrk 509.
To generate these simulated light curves, we follow the procedure
described by \citet{Peterson98} and \citet{Horne04}.
We first construct a power density spectrum with a spectral index randomly
drawn from a Gaussian distribution with a mean $\alpha = 2.0$ and a dispersion
of 0.5.
Since the power density spectrum is the Fourier pair of the autocorrelation
function, taking the square root of this distribution then gives the Fourier
amplitudes of the light curve. As described by Peterson et al. (1998), a
random lightcurve can then be generated by assigning random phases to these
amplitudes and then taking the inverse Fourier transform.
To normalize the mean flux and fractional variations in this light curve,
we use the historical UV data for Mrk 509 compiled by \citet[see our
Figure~\ref{fig:lc}]{Dunn06},
updated with our new COS observation.
For these data, binned to 200-day timescales, we measure a mean flux at
1401 \AA\ of $7.04 \times 10^{-14}$~ergs~cm$^{-2}$~s$^{-1}$~\AA$^{-1}$ and a
fractional variation $F_{var} = 0.29$ (where $F_{var}$ is as defined by
\citealt{Rodriguez97}).

We solve the coupled time-dependent differential equations
(Equation~\ref{eq:ionfrac}) for a given element numerically using the 4th order
Runge-Kutta method. The initial recombination coefficients ($\alpha_X(T)$) and
column densities (see Table~\ref{table:iv}) are taken from the best-fit {\sc
Cloudy} models with parameters given in Table~\ref{table:models} for each
trough. We compute ionization rates using Equation~\ref{eq:steadystate}. We do
not use the ionization rates provided by {\sc Cloudy} since those rates do not
result in equilibrium using the simplified formalism that leads to
Equation~\ref{eq:steadystate} (see also Paper VIII). In previous papers
(\citealt{Mehdipour11}, hereafter Paper IV; Paper VIII), we have shown that flux
variability in the optical, ultraviolet, and soft X-rays in Mrk 509 is highly
correlated, which gives us confidence that the portion of the SED most important
for the ionization of \civ\ and \nv\ maintains a constant shape even as the
overall normalization varies. We therefore assume that the SED maintains a
constant shape for the entire time period. The simulated lightcurves, discussed
above, extend over a period of 22 years. From those lightcurves, we select the
ones that have a flux value at $t_2$ that is approximately 70\% higher
than that at $t_1$, where $t_2-t_1 = 8$~yr (the time between the STIS and COS
epochs). We use the simulated lightcurve only in the interval $t_1 < t < t_2$ in
order to match the measured flux levels in the 2001 and 2009 epochs. From the
1000 original lightcurves, 928 contain regions that fit these criteria. Since 7
of the 928 lightcurves have two 8~yr periods separated by at least 6 months that
fit our criteria, we have a sample of 935 lightcurves.

UV flux monitoring of the 100 days before the COS observations (Paper IV,
Figure 2) reveals that the quasar continuum over that time interval was always
at least 70\% above that during the 2001 STIS epoch (Fig. 3, bottom panel). We
therefore fix the last 100 days of all the lightcurves in our sample to be constant at $\simeq 1.72$ times the STIS value as a conservative estimate for
the flux change.

For our initial conditions at the STIS 2001 epoch, we assume that the absorber
was in photoionization equilibrium at that time. As can be seen in the top
panel of Fig.~\ref{fig:lc}, there is gap of several years between the STIS
observation (the last point on the plot) and the previous IUE monitoring. We
therefore have limited information about the lightcurve behavior prior to the
2001 epoch. However, both Fig.~\ref{fig:lc} (center panel) and our 2009 UV
monitoring (bottom panel) suggest that the Mrk 509 UV flux changes gradually over timescales of
50-100 days. In the center panel of Fig.~\ref{fig:lc}, the flux varies by a
maximum of $\sim 40$\%, while our 2009 UV monitoring reveals maximum flux
changes of $\sim 30$\% (also see Paper IV, Figure 2).
Therefore, it is plausible that the UV flux in the 100 or so days before the
STIS epoch was similar to that of the actual measurement during the 2001
observing epoch. Moreover, we note that
FUSE observations in 1999 and 2000 show that the flux at 1175\AA\ was within
~10\% of that for the STIS observations (see Table 2 in
Paper VI). These two additional lightcurve points suggest that the low flux
state of Mrk 509 probably existed in the two years prior to the 2001 STIS epoch.
Under this assumption, as long as the recombination timescale is shorter than 2
years, we can use the photoionization equilibrium assumption. For lower density
plasma ($n_e\ltsim10^3$, see table 5) the plasma cannot be approximated as being
in photoionization equilibrium even if the flux was constant for the previous
two years. However, $n_e \simeq 10^3$ is roughly the upper limit we obtain from
the full time-dependent solution for most components (see
Table~\ref{table:densities}). A plasma with a lower $n_e$ will be at a larger
distance than the lower limits we derive in this paper and thereby, consistent
with our results.


\begin{figure}
  \includegraphics[angle=90,width=1.0\hsize]{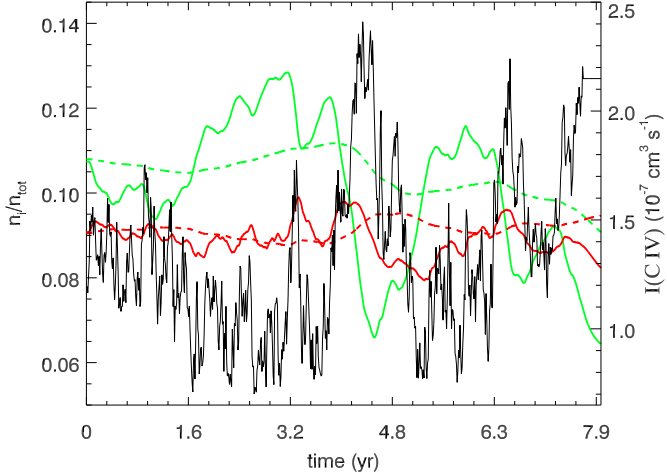}\\
 \caption{Simulations of \textsc{Civ}/C (green) and \textsc{Nv}/N (red) versus
time for one lightcurve in the sample. The solid lines are for a density
$\vy{n}{H} = 10^{3.5}$ and the dashed lines are for $\vy{n}{H} = 10^{2.5}$. The
black line is the ionization rate per particle for \civ\ and is propotional to
the simulated lightcurve for this example.}
 \label{simulation} 
\end{figure}

\begin{figure*}
  \includegraphics[width=0.5\textwidth]{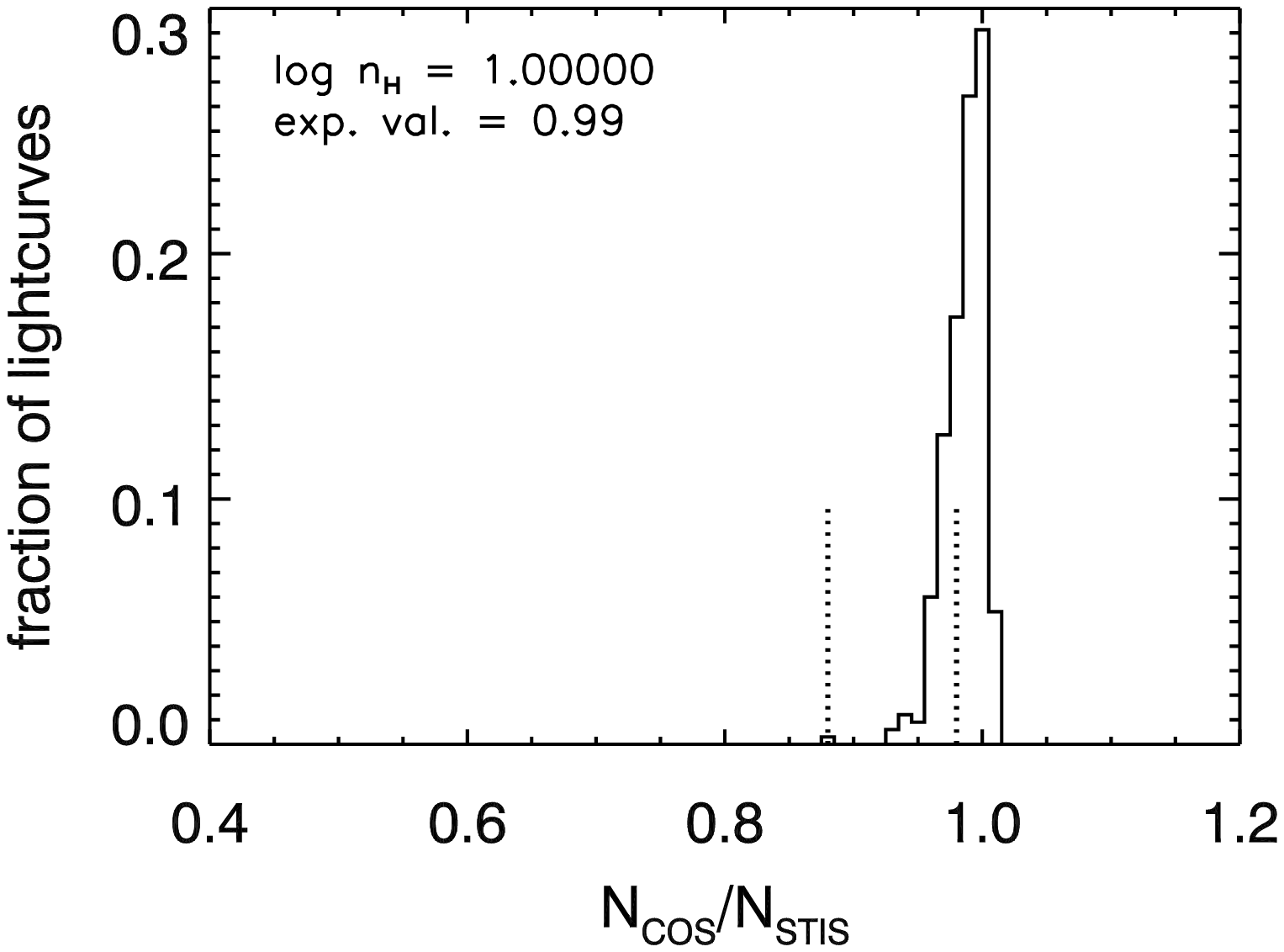}
  \includegraphics[width=0.5\textwidth]{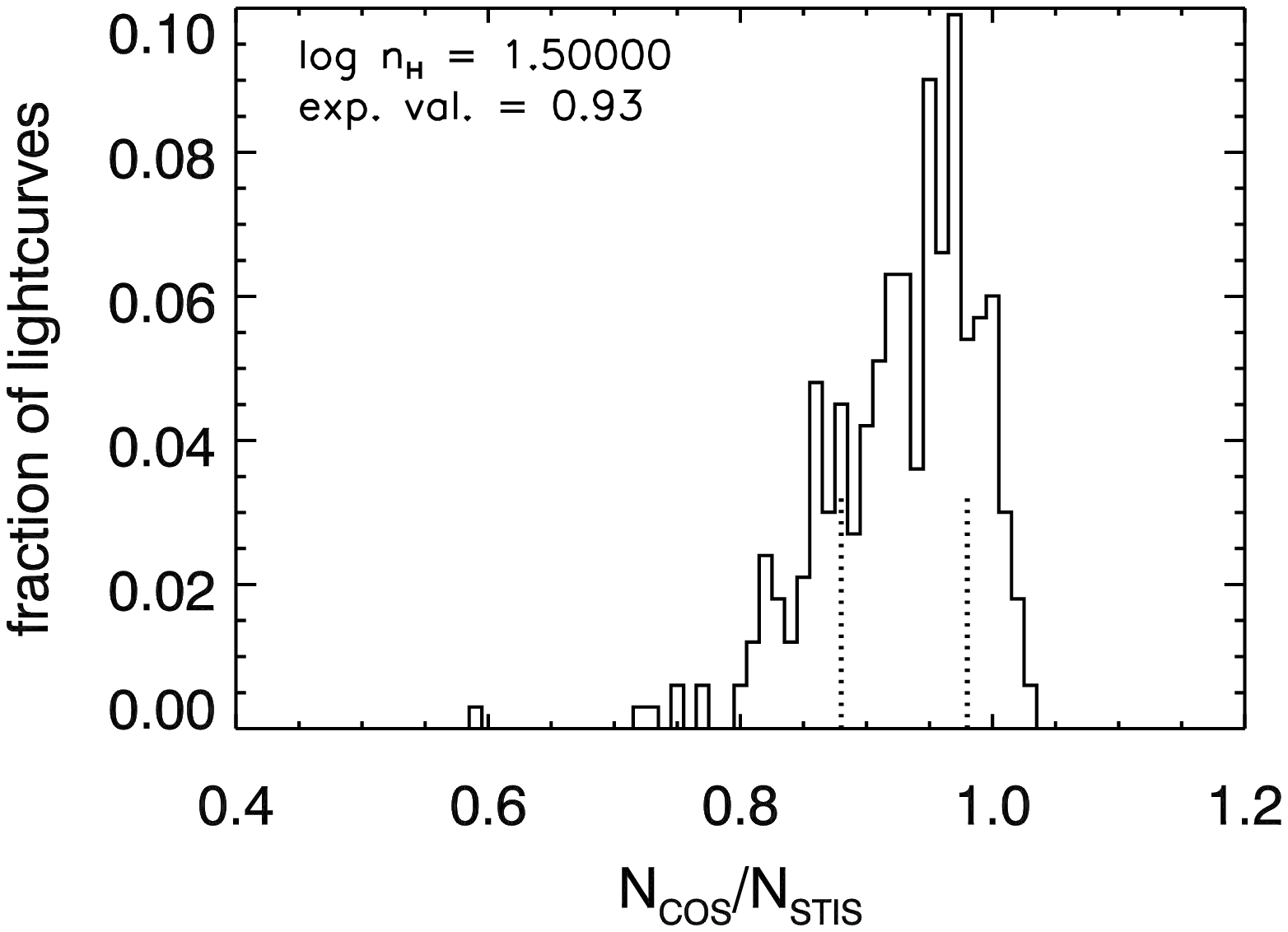}
  \includegraphics[width=0.5\textwidth]{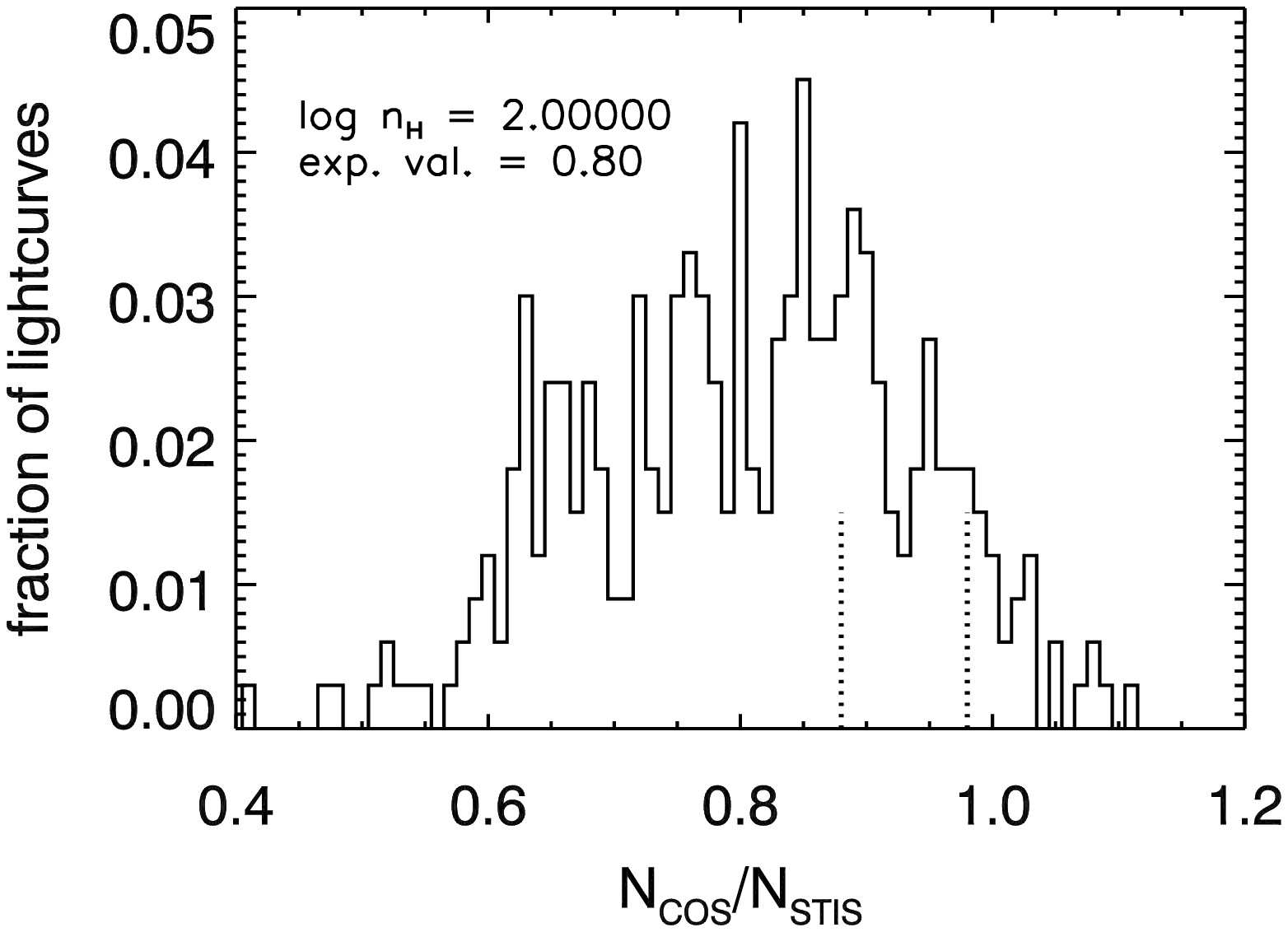}
  \includegraphics[width=0.5\textwidth]{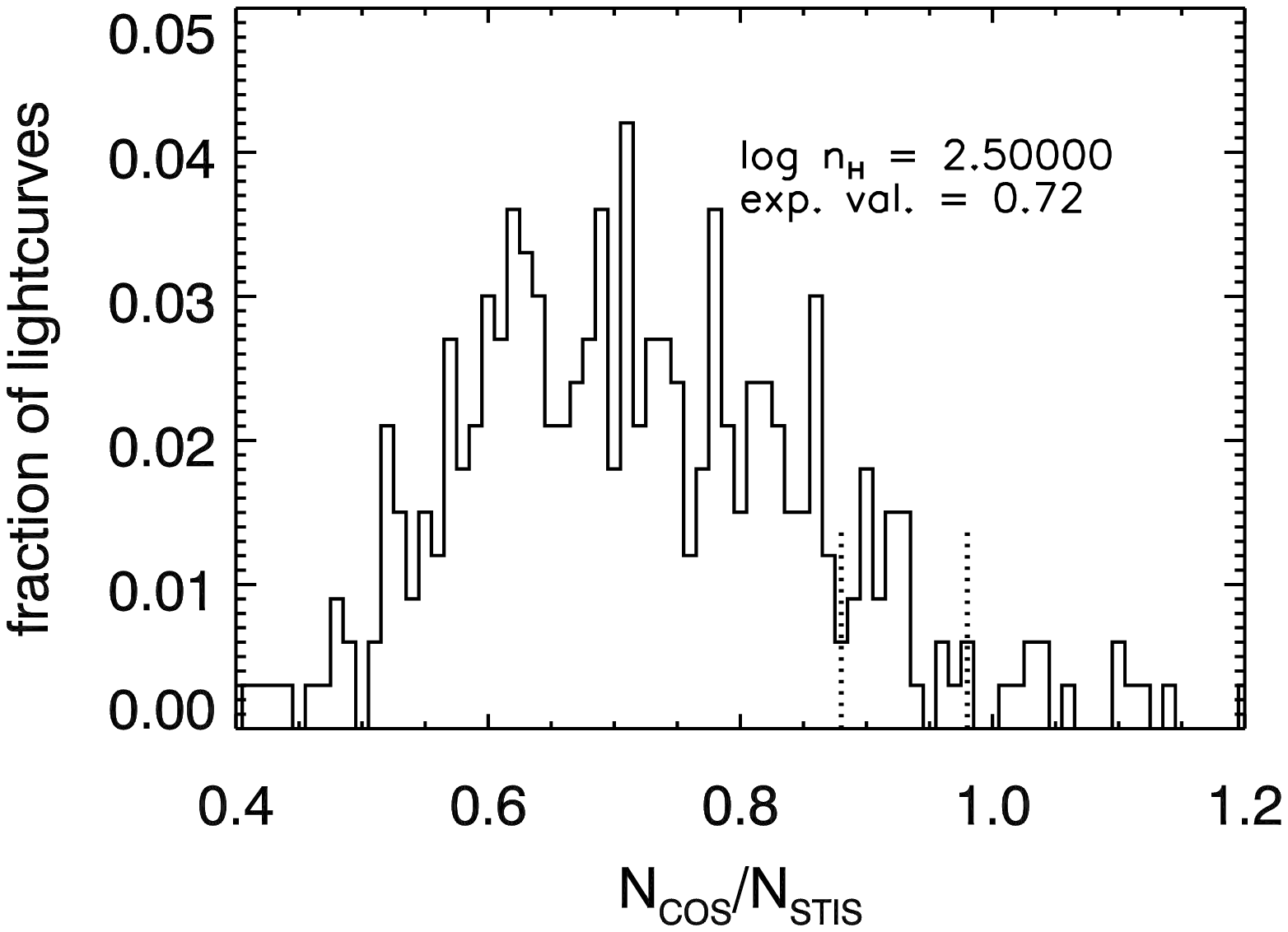}
  \includegraphics[width=0.5\textwidth]{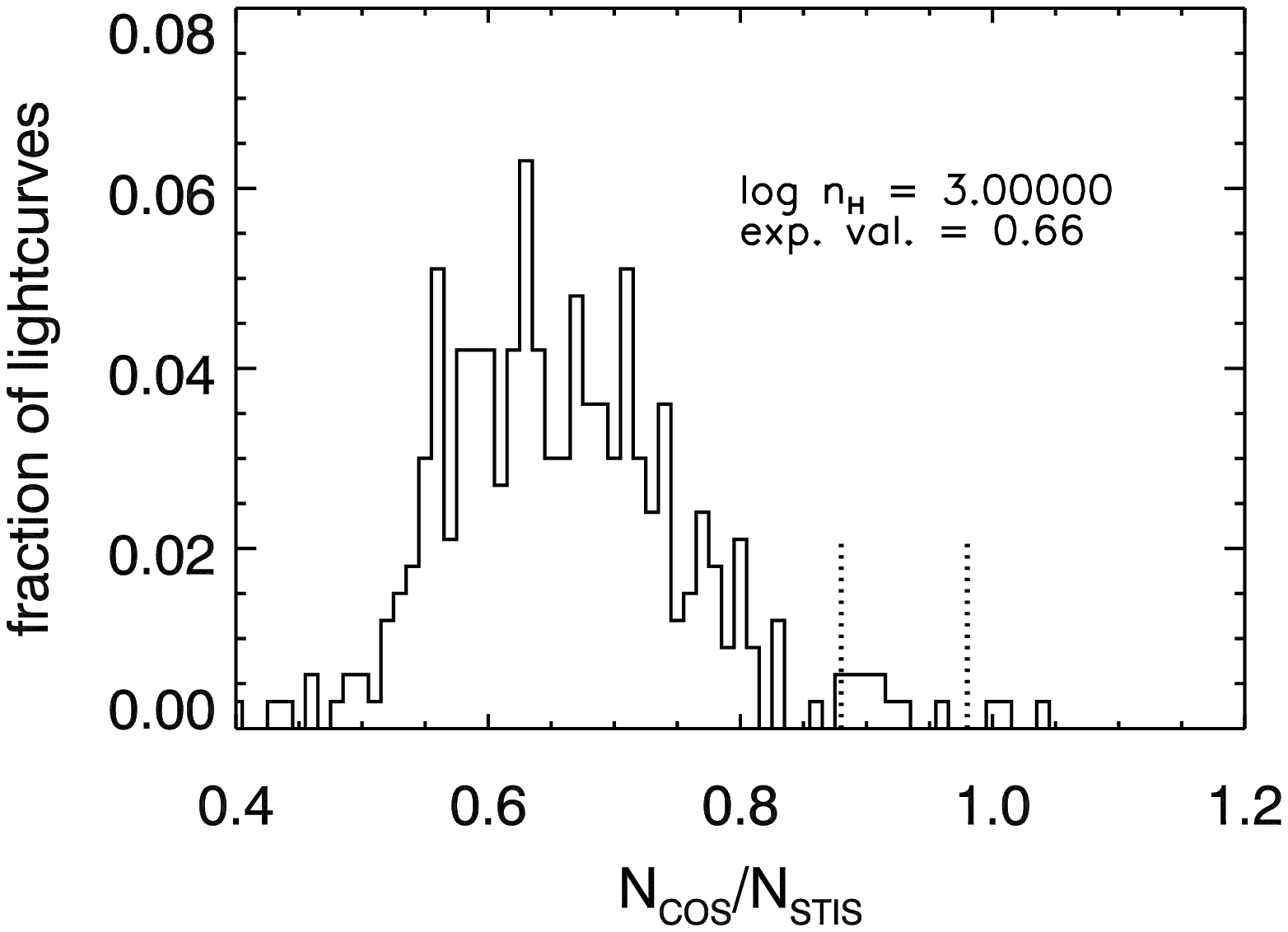}
  \includegraphics[width=0.5\textwidth]{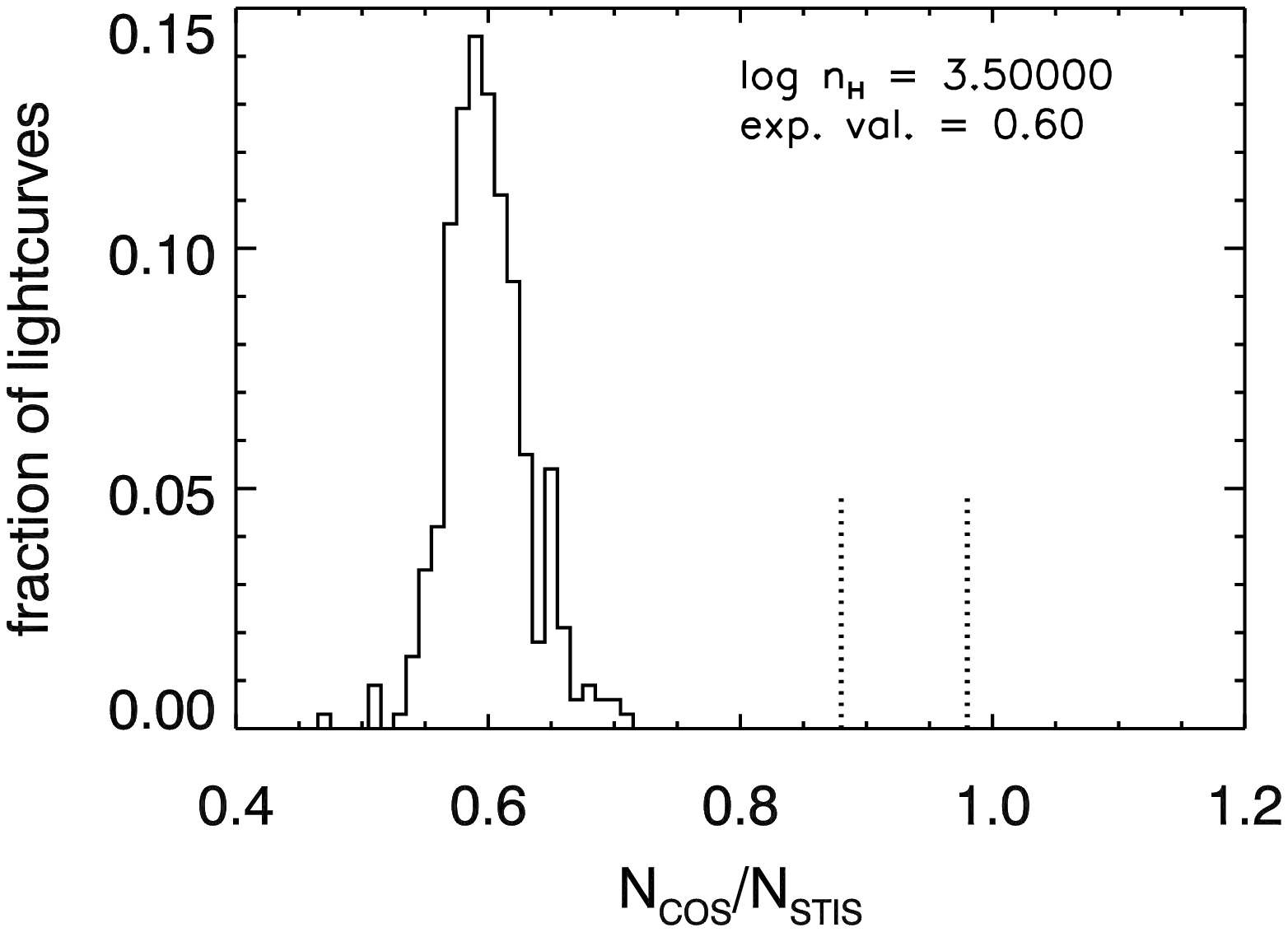} \\
 \caption{Determining the upper limit for $\vy{n}{H}$ of the absorber. This
example shows the results for trough T2, based on the \civ\ column density
measurements.  For each of the 935 simulated light curves (see section 5), we
explicitly solve the time-dependent photoionization equations and calculate the
expected ratio of \civ\ column density over a time period equal to the
differences between the STIS and COS epochs. Each panel represents the
fractional solution distribution of $N$(\civ)$_{COS} / N$(\civ)$_{STIS}$ for a
given $\log(\vy{n}{H})$ (cm$^{-3}$), which increases in the panels from left to
right and top to bottom. The vertical dashed lines  show the positive and
negative 1$\sigma$ errors on the measured value. We determine an $\vy{n}{H}$
upper limit by increasing $\vy{n}{H}$ until more than 99\% of the calculated
$N$(\civ)$_{COS} / N$(\civ)$_{STIS}$ are outside the error bars. In this
example, we find $\vy{n}{H} < 10^{3.2}$ cm$^{-3}$.}
 \label{histograms} 
\end{figure*}

To determine an upper limit on the electron number density, we simulate the
time-dependent changes in column densities of \civ\ and \nv\ and compare them
with the limits imposed by the observed differences between the STIS and COS
data (see Table~\ref{line_data}). For each given $n_e$, we track the fractional
change in $N$(\civ) and $N$(\nv) for all 935 simulated lightcurves in each of
the seven troughs
for which we have an initial photoionization solution. In Fig.~\ref{simulation},
simulations for \civ\ and \nv\ for two different electron number densities are
shown for one of the simulated lightcurves with $t=0$ corresponding to the STIS
epoch and $t \simeq 8$~yr corresponding to the COS epoch.
From the simulations, histograms of the fraction of simulations versus the
predicted change in column density are produced for each ion in each trough.
We choose the upper limit on $n_e$ as the lowest density for which more than
99\% of the lightcurves predict changes greater than those suggested by the
data. Figure~\ref{histograms} shows an example of the resulting histograms, and
Table \ref{table:densities} lists the results for each trough and ion.

Of components T1--T7, only T1 and T2 show a significant change in column
densities. \civ\ is saturated in trough T1 and shows no significant change, but
\nv\ shows $>3\sigma$ changes in residual intensity for both components of the
doublet. In trough T2, change is observed for \civ, while no change is observed for \nv. Since these two components have responded to continuum changes, we can put a
lower limit on $n_e$ and thereby, an upper limit on the distance. We do this by
finding the highest density for which more than 99\% of the lightcurves predict
changes smaller than those suggested by the data. With distances $\ltsim
2.1$~kpc, these absorbers are within the confines of the host galaxy. We are also
able to put an upper limit on $n_e$ using the same method. We note that our simulations for trough T1 predict changes that are \textit{smaller} than that
measured for both high and low densities. This is because the ionization
parameter is near the value producing the highest \nv\ fractional abundance
($U_H = 10^{-1.5}$). As we increase $n_e$ from $10$~cm$^{-3}$ to
$10^4$~cm$^{-3}$, the ionization state of the gas at the COS epoch increases.
For $n_e \ltsim 100$~cm$^{-3}$, the change in $n$(\nv) between STIS and COS
epochs increases with increasing density. However, for $n_e \gtsim
100$~cm$^{-3}$, the \nv\ column density decreases as the ionization of the gas
becomes higher than than that producing the highest fractional abundance of \nv.
At densities of $n_e \simeq 10^4$~cm$^{-3}$, the lower ionization state at the
time of the STIS observation and the higher ionization state at the time of the
COS observations produce approximately the same amount of \nv, and it therefore
appears as if there is no change between epochs. For even higher densities,
simulations predict a decrease in $n$(\nv) between the STIS and COS epochs.

\begin{table}[!tbp]
\caption{density and distance limits from simulations} 
\centering                
\begin{tabular}{c@{\,}c@{\,}c@{\,}c@{\,}cc} 
\hline\hline
Trough & $v$ & log $n_e$(\textsc{Civ}) & log $n_e$(\textsc{Nv}) &
$R_{99}$\tablefootmark{a} & $R_{90}$\tablefootmark{b} \\
 & (km s$^{-1}$) & (cm$^{-3}$) & (cm$^{-3}$) & (pc) & (pc) \\
\hline
 T1 & -405 & $\cdots$\tablefootmark{c} & 1.1--4.2 & 60--2100 & 80--1500 \\
 T2 & -310 & 1.1--3.2 & $<$4.2 & 160--1830 & 370--1460 \\
 T3 & -240 & $<$3.2 & $<$3.9 & $>$130 & $>$290  \\
 T4 & -70 & $<$3.1 & $<$3.4 & $>$130 & $>$290  \\
 T5 & -15 & $<$3.6 & $\cdots$\tablefootmark{d} & $>$130 & $>$260  \\
 T6 &  +45 & $<$2.8 & $<$3.2 & $>$150  & $>$370 \\
 T7 &  +125 & $<$2.5 &  $\cdots$\tablefootmark{e} & $>$130 & $>$290  \\
\hline
\label{table:densities}                         
\end{tabular}
\tablefoot{
\tablefoottext{a}{Distances determined by requiring $> 99$\% of the lightcurves
to overpredict changes in column density.}
\tablefoottext{b}{Distances determined by requiring $> 90$\% of the lightcurves
to overpredict changes in column density.}
\tablefoottext{c}{Since \civ\ is saturated, we have no information about changes
in $N$(\textsc{Civ}). }
\tablefoottext{d}{For T5, the change predicted for the column density of
\nv\ is within the error regardless of number density precluding the
determination of a useful limit.}
\tablefoottext{e}{\nv\ in trough T7 shows change in the red component but not
the blue component yielding contradictory results.}
}\end{table}

We use the upper limits placed on the hydrogen number density $\vy{n}{H}$
(recall that $\vy{n}{H} \simeq n_e/1.2$ in highly ionized plasma) to determine
lower limits on the distance to the absorber from the ionizing source via the
ionization parameter (Equation~\ref{eqionparam}). These are given in
Table~\ref{table:densities}, where $R_{99}$ and $R_{90}$ are the distances
determined by requiring that 99\% and 90\% of the lightcurves, respectively,
give results inconsistent with the differences in measured values. Except for
component T1, these distances are determined using the $n_e$ derived from \civ\
since they give the smallest upper limit consistent with both the \civ\ and \nv\
simulations. The rate of ionizing photons striking the gas is determined by
fitting our SED to the 1175\AA\ flux from STIS data given in Paper VI. We find
$Q_H = 5.03 \times 10^{54}$~s$^{-1}$.


\section{Discussion}

We were able to put conservative lower limits on the distance to the
absorber of 100--200 pc from the ionizing source for all the UV components using
the fact that the column
densities of \civ\ and \nv\ showed little or no variation between the
STIS and COS epochs despite a large change in ionizing flux.
Since the  lightcurve for Mrk 509 was not densely monitored between the epochs,
we used Monte Carlo simulated lightcurves to statistically determine the
distance limits.

The limits on the distance computed using timescale arguments (see Section 4) are similar to and bracket those we found statistically. For the lightcurve that increased just after the STIS epoch, the distance limits are 2.0 to 3.2 times larger
than our 99\% simulation based limits. For the lightcurve that increased just before the 100 days monitoring before the COS epoch, the distance limits are 1.6 to 2.6 times smaller than our 99\% simulation based limits. Using the second lightcurve in the time-dependent ionization equations and requiring changes in ionic column density to be smaller than the 1-$\sigma$ errors given in Table~2 yields distances that are similar to our 90\% simulation based limits and within a factor of 2 larger than our 99\% simulation based limits for all troughs except T6 (factor of $\sim 3$ larger).

Our distance results are consistent with those derived for the 
simultaneous X-ray absorber data.
In Paper III, five discrete ionization components were identified in
the \textit{XMM-Newton} spectrum of Mrk~509, named A, B, C, D, and E.
Our analysis of the lack of spectral variability of these X-ray components
during our campaign combined with variations seen in comparison with archival
data (Paper VIII) showed
that component C has a distance of $>$70 pc, component D is between 
5 and 33 pc, and component E has a distance between 5 and 21--400 pc,
depending upon modeling details. For the lowest ionization components, A and B,
we were not able to establish
any significant limits on the gas density or the distance.
These low-ionization components, however, are closely associated with
the UV components, so the bounds on distance that we establish in this paper
completes our overall picture of the outflow in Mrk 509.

Based on the 100--200 pc lower limit for all the UV components, this 
absorber cannot be connected with an accretion disc wind.
The outflow might have originated 
from the disc, but based on simple ballistic 
kinematics, such an event had to occur at least 300,000 years ago in the 
rest frame of the source.
\vspace{1cm}

\citet{Phillips83} found extended emission in an area 6.6 kpc in diameter
centered on the nucleus of Mrk 509. The radial velocities they measured for
their high-ionization component correspond to the velocities for our troughs
T2--T5, indicating that we may be seeing the same outflow. They also find a
low-ionization component with line intensity ratios similar to Galactic \hii\
regions and velocities corresponding to our troughs T3--T7. If we are seeing the
same outflow in absorption features as \citet{Phillips83} saw in emission
features, the distance to the absorbers is $\gtsim 3$~kpc (assuming a conical
outlflow with an opening angle of $\ltsim 45^\circ$), putting them on scale with
galactic winds \citep{Veilleux05}.

\begin{acknowledgements}
This work is based on observations obtained with the \textit{Hubble} Space Telescope (HST), a cooperative program of ESA and NASA. Support for HST Program number 12022 was provided by NASA through grants from
the Space Telescope Science Institute, which is operated by the Association of
Universities for Research in Astronomy, Inc., under NASA contract NAS5-26555. We also made use of observations obtained with \textit{XMM-Newton}, an ESA science mission with instruments and contributions directly funded by ESA Member States and the USA (NASA), as well as data supplied by the UK \textit{Swift} Science Data Centre at the University if Leicester.
SRON is supported financially by NWO, the
Netherlands Organization for Scientific Research. J.S. Kaastra thanks the PI of
\textit{Swift}, Neil Gehrels, for approving the TOO observations. M. Mehdipour
acknowledges the support of a PhD studentship awarded by the UK Science \&
Technology Facilities Council (STFC). N. Arav and G. Kriss gratefully
acknowledge support from NASA/\textit{XMM-Newton} Guest Investigator grant NNX09AR01G. D.
Edmonds and B. Borguet were supported by NSF grant 0837880. E. Behar was
supported by a grant from the ISF. S. Bianchi, M. Cappi, and G. Ponti
acknowledge financial support from contract ASI-INAF n. I/088/06/0. P.-O.
Petrucci acknowledges financial support from CNES and the French GDR PCHE. G.
Ponti acknowledges support via an EU Marie Curie Intra-European Fellowship under
contract no. FP7-PEOPLE-2009-IEF-254279. K. Steenbrugge acknowledges the support
of Comit\'e Mixto ESO - Gobierno de Chile. Finally, we would like to thank the referee for many useful comments.

\end{acknowledgements}

\appendix

\section{Behavior of the Time-dependent Photoionization Equations}

As an illustrative example we look at the simple case of hydrogen.
We are interested in the changes of neutral hydrogen in response to changes in
ionizing flux.
From equation (\ref{eq:ionfrac}) we obtain
\begin{equation}
\frac{d\vy{n}{H0}}{dt}=-\vy{n}{H0}I_{H0}+\vy{n}{H^+}R_{H0}.
\label{eq:H0frac}
\end{equation}
Let us assume that we start from a steady state ionization equilibrium
(Equation~\ref{eq:steadystate}) with $I_{H0}=I_0$
, and that at $t=0$ the absorber
experiences an instantaneous flux change: $I_{H0}(t>0)=I_0(1+f)$, where $f$ can
be either positive or negative. We therefore obtain
\begin{equation}
\frac{d\vy{n}{H0}}{dt}=-(1+f)\vy{n}{H0}I_0+\vy{n}{H^+}R_{H0}
\label{eq:H0frac2}
\end{equation}
Assuming
$\vy{n}{H^+}/\vy{n}{H0}\gg 1$ (as is typical for AGN outflow material), for an
order of magnitude increase or decrease in flux, $\vy{n}{H^+}$ stays constant to
a high degree,
and therefore the right most term in equation (\ref{eq:H0frac2}) can be treated
as constant.
Under these assumptions there is a simple analytical solution for equation
(\ref{eq:H0frac2}):
\begin{equation}
\vy{n}{H0}(t)=\vy{n}{H0}(0)\frac{1+fe^{-(1+f)I_0 t}}{1+f}
\label{eq:H0fracSOL}
\end{equation}
This solution satisfies the differential equation as well as the two boundary
conditions: $\vy{n}{H0}(t=0)=\vy{n}{H0}(0)$
and $\vy{n}{H0}(t=\infty)=\vy{n}{H0}(0)/(1+f)$, where the latter condition stems
from the new steady state reached with flux level
$I_0(1+f)$ (see Equation~\ref{eq:steadystate})
Several properties of this solution are worth mentioning:

\begin{enumerate}
 \item If we start from an ionization equilibrium, the  timescale for changes in
$\vy{n}{H0}(t)$ is
\begin{equation}
 [(1+f)I_0]^{-1}=[(1+f)\vy{\alpha}{H^+}n_e(\vy{n}{H^+}/\vy{n}{H0})_{t=0}]^{-1},
\end{equation}
which is inversely proportional to $n_e$. This is the timescale for 63\%
(1-$e^{-1}$) of the total $\Delta \vy{n}{H0} \equiv
\vy{n}{H0}(0)-\vy{n}{H0}(\infty)$ change to occur. For $|f| \ll 1$, this
timescale is approximately equal to that given by Equation~\ref{eq:trec}.
\item For $f\gg1$ the timescale for changes in $\vy{n}{H0}(t)$ is roughly given
by $(I_0f)^{-1}$.
\item  For a situation where $I_{H0}(t>t_0)$ drops instantaneously back to $I_0
$ we obtain two interesting limits: \\
a) when $t_0\gg [(1+f)I_0]^{-1}$, $\vy{n}{H0}(t_0)=\vy{n}{H0}(0)/(1+f)$, i.e.,
ionization equilibrium has been reached. \\ b) when $t_0\ll [(1+f)I_0]^{-1}$,
$\vy{n}{H0}(t_0)=\vy{n}{H0}(0)(1-fI_0t_0)$, which represent a damping in the
maximum variation of $\vy{n}{H0}(t)$ that is inversely proportional to $n_e$ for
a given $t_0$
\end{enumerate}

\begin{figure}
  \includegraphics[width=1.0\hsize]{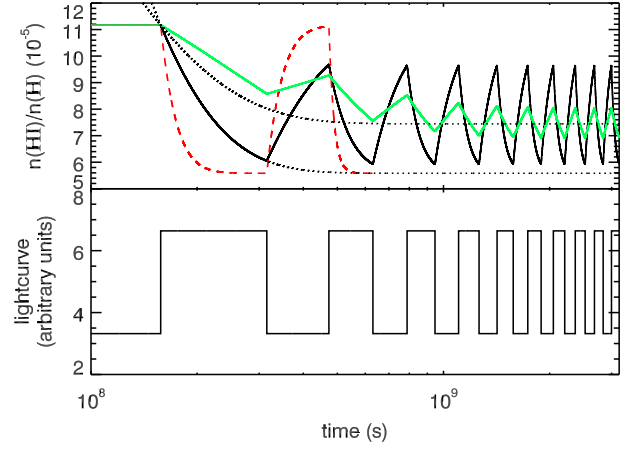}\\
 \caption{In the upper panel, simulations of the fraction of H in\textsc{Hi} are
plotted for a periodic step function light-curve with $\Delta t = 1.6 \times
10^8$~s (shown in the lower panel). The hydrogen number density of the thick
black line was chosen to give a recombination timescale equal to half the period
of the lightcurve. The red dashed line has a timescale that is 4 times shorter,
and the green line has a timescale that is 4 times longer. The dotted black
lines are analytic solutions for a flux change f=1 (lower line) and f=0.5 (upper
line).} 
 \label{hsim}
\end{figure}

The above three properties give insight to more physically interesting scenarios
such as the ionization behavior of the absorber to abrupt cyclical changes in
ionizing flux (see Figure~\ref{hsim}).

Assuming $I_{H0}(t<0)=I_0$ and that $n_e$ is roughly constant at all times, we
start with an ionization equilibrium
$(\vy{n}{H^+}/\vy{n}{H0})_{t<0}=(I_{H0}/R_{H0})_{t<0}$.
Following the three points above we expect that for $t_0\gg
[(1+f)\vy{\alpha}{H^+}n_e(\vy{n}{H^+}/\vy{n}{H0})_{t=0}]^{-1}$
the absorber will quickly oscillate between the equilibria values:
$\vy{n}{H0}=\vy{n}{H0}(0)/(1+f)$ and $\vy{n}{H0}(0)$.
That is, in this limit the plasma has no memory for the history of ionizing flux
changes and $\vy{n}{H0}$ closely follows the
current value of $I_{H0}$. The situation is different for $t_0\ll
[(1+f)\vy{\alpha}{H^+}n_e(\vy{n}{H^+}/\vy{n}{H0})_{t=0}]^{-1}$. Initially, each
$t_0$ period of enhanced flux decreases
$\vy{n}{H0}$ by a factor $fI_0t_0$, where $I_0t_0\ll 1$ by definition. After
many cycles ($N\gg(I_0t_0)^{-1}$) a pseudo equilibrium is reached where
$\vy{n}{H0}=\vy{n}{H0}(0)/(1+f/2)$. In this case the plasma has a strong memory
for the history of ionizing flux changes and the psuedo equilibrium depends on
the average ionizing flux over $\Delta t\sim I_0^{-1}$.

For elements other than hydrogen, it is not possible to obtain useful analytical
solutions for equation set (\ref{eq:ionfrac}).
However the qualitative behavior is quite similar. \\

\bibliographystyle{aa}
\bibliography{astro}

\end{document}